\documentclass[%
 reprint,
 amsmath,amssymb,
 aps,
]{revtex4-2}

\usepackage{graphicx}
\usepackage{dcolumn}
\usepackage{bm}
\usepackage[colorlinks=true,citecolor=blue]{hyperref}
\usepackage[utf8]{inputenc}

\usepackage[mathlines]{lineno}

\usepackage{aas_macros}

\begin{document}

\preprint{}

\title{Early warning of coalescing neutron-star and neutron-star-black-hole binaries from nonstationary noise background using neural networks}

\author{Hang Yu}
\email{hangyu@caltech.edu}
 \affiliation{TAPIR, Walter Burke Institute for Theoretical Physics, MC 350-17 California Institute of Technology, Pasadena, CA 91125, USA}

\author{Rana X. Adhikari}%
\affiliation{%
LIGO Laboratory, California Institute of Technology, MC 100-36, Pasadena, CA 91125, USA}%

\author{Surabhi Sachdev}
\affiliation{%
Department of Physics, The Pennsylvania State University, University Park, PA 16802, USA}
\affiliation{Institute for Gravitation and the Cosmos, The Pennsylvania State University, University Park, PA 16802, USA }

\author{Ryan Magee}
\affiliation{%
LIGO Laboratory, California Institute of Technology, MC 100-36, Pasadena, CA 91125, USA}

\author{Yanbei Chen}
\affiliation{TAPIR, Walter Burke Institute for Theoretical Physics, Mailcode 350-17 California Institute of Technology, Pasadena, CA 91125, USA}

\date{\today}

\begin{abstract}
The success of the multi-messenger astronomy relies on gravitational-wave observatories like LIGO and Virgo to provide prompt warning of merger events involving neutron stars (including both binary neutron stars and neutron-star-black-holes), which further depends critically on the low-frequency sensitivity of LIGO as a typical binary neutron star stays in this band for minutes. However, the current sub-60 Hz sensitivity of LIGO has not yet reached its design target and the excess noise can be more than an order of magnitude below 20\,Hz. It is limited by nonlinearly coupled noises from auxiliary control loops which are also nonstationary, posing challenges to realistic early-warning pipelines.
Nevertheless, machine-learning-based neural networks provide ways to simultaneously improve the low-frequency sensitivity and mitigate its nonstationarity, and detect the real-time gravitational-wave signal with a very short computational time. We propose to achieve this by inputting both the main gravitational-wave readout and key auxiliary witnesses to a compound neural network. Using simulated data with characteristic representing the real LIGO detectors, our machine-learning-based neural networks can reduce nonlinearly coupled noise by about a factor of 5 and allows a typical binary neutron star (neutron-star-black-hole) to be detected 100 s (10 s) before the merger at a distance of 40 Mpc (160 Mpc). If one can further reduce the noise to the fundamental limit, our neural networks can achieve detection out to a distance of 80 Mpc and 240 Mpc for binary neutron stars and neutron-star-black-holes, respectively. It thus demonstrates that utilizing machine-learning-based neural networks is a promising direction for the timely detection of the coalescence of electromagnetically bright LIGO/Virgo sources.
\end{abstract}

\maketitle


\section{Introduction}

The current generation of ground-based gravitational-wave interferometers~\cite{TheLIGOScientific:2014jea,TheVirgo:2014hva,Akutsu:2018axf} firmly established a new way to observe our cosmos. Since the first detection of gravitational waves (GWs) from a binary black hole (BBH) merger~\cite{Abbott:2016blz}, Advanced LIGO (aLIGO~\cite{TheLIGOScientific:2014jea}) and Advanced Virgo (aVirgo~\cite{TheVirgo:2014hva}) have gone on to document dozens of gravitational-wave candidates~\cite{LIGOScientific:2018mvr,Abbott:2020niy} that have been confirmed and added to by the broader astrophysical community~\cite{Nitz:2018imz,Nitz:2019hdf,Zackay:2018qdy,Zackay:2019kkv,Zackay:2019tzo,Venumadhav:2019tad,Venumadhav:2019lyq,Magee:2019vmb}.
One of the most spectacular discoveries made by Advanced LIGO and Virgo is the first observed binary neutron star (BNS) coalescence, GW170817. GW170817 was jointly detected in low-latency in gravitational waves~\cite{GW170817a} and by Fermi-GBM in gamma rays~\cite{Goldstein:17}. The subsequent discovery and followup of kilonova AT 2017gfo led to a concerted followup effort across the electromagnetic (EM) spectrum~\cite{GW170817b}. The resulting multi-messenger observations enabled an abundance of new science:
constraints on the maximum NS mass~\cite{LIGOScientific:2019eut}, 
better understandings of neutron star mode coupling and equation of state~\cite{Weinberg:2018icl,Abbott:2018exr,Radice:2017lry},
as well as tests of general relativity~\cite{Abbott:2018lct}. 

Despite the successes surrounding GW170817, there is still much to be learned about compact binary mergers containing at least one neutron star. In particular, there are various astrophysical processes that can generate precursor and/or early-stage signals that are yet to be detected. For example, tidal interactions might shatter crusts of neutron stars and lead to short gamma-ray burst~\cite{Tsang:12}. The property of the final merger product may be better revealed with prompt X-ray and optical observations~\cite{Metzger:14}. In the radio band, precursor magnetosphere interactions might cause radio emissions~\cite{Hansen:01, Most:20} and could be a potential mechanism leading to fast radio
bursts~\cite{Thornton:13, Totani:13}. See, e.g., Ref.~\cite{Sachdev:20} for further discussions on potential early-warning signals as well as a nice summary of the follow-up capacity of various  EM observatories.


To detect the prompt signatures of these processes, LIGO and Virgo would need to be able to identify the existence of a GW event and then determine its sky location in a timely manner. This is especially important for binaries where at least one component is a neutron star. The GW alert for GW170817 was not sent out until $\sim 40$ minutes after the merger and the sky location was not released until another 4 hours later~\cite{GW170817b};
in principle, this information can be obtained minutes \emph{prior to} the final merger as a typical BNS event will stay in the sensitivity band of LIGO and Virgo for minutes at their designed sensitivities. 

There are presently four low-latency, matched-filter based pipelines that produce near real-time gravitational-wave alerts for BNS and NSBH mergers: \texttt{GstLAL}~\cite{Cannon:12, Messick:17}, \texttt{PyCBC}~\cite{Nitz:18}, \texttt{MBTA}~\cite{Adams:16}, and \texttt{SPIIR}~\cite{Luan:12, Chu:20}. Several of these pipelines have already developed analyses capable of early warning detection~\cite{Cannon:12,Sachdev:20,Chu:20}.
See also Ref.~\cite{LSC:19} for a summary of current efforts carried out by the LIGO and Virgo collaborations during the second observing run to low-latency warnings. 
The prospect of pre-merger detection is ultimately limited by latencies surrounding data acquisition, handling, and analysis. Ref.~\cite{2021arXiv210204555M} recently demonstrated that even at present latencies, the LIGO-Virgo collaboration is capable of identifying, localizing, and broadcasting GW candidates prior to merger.

Machine-learning (ML) based neural networks (NNs) offers yet another attractive alternative to achieve the early warning of BNSs/NSBHs. Instead of individually computing the overlap between a time series of GW readout and each waveform template from a large template bank, a trained NN would only need to do the computation once to predict the existence and the property of the source. It can therefore serve as the first step for existing pipelines and further accelerate their computational efficiency. 

Indeed, various authors have considered the possibility of detecting GW events using ML-based NNs. Refs.~\cite{George:18a, George:18b, Gabbard:18} showed that it is possible to input real-time GW readout and then use NNs to detect massive black hole binaries (BBHs) and later Refs.~\cite{Lin:20, Krastev:20} considered the possibility of detecting BNSs with longer signal duration. Recently, Ref.~\cite{Baltus:21} further considered detecting BNS events tens of seconds prior to the final merger. 

However, almost all the analyses above assume a stationary Gaussian noise background and often at the designed sensitivity of aLIGO (one exception is Ref.~\cite{George:18b}, yet they focused on short BBH signals only with duration $\lesssim 1\,{\rm s}$ long in which the nonstationarity is less critical). While this is a decent approximation in the $f>100\,{\rm Hz}$ frequency band, at lower frequencies which matter most for the early warning the dector noise not only exceeds the designed level by orders of magnitude but also exhibits nonstationarity~\cite{Martynov:16, Zackay:2019kkv, Buikema:20}. Therefore, it would be crucial to take into account these features of realistic detector noise in order to design a NN to achieve early warning in practice. 

Our work thus extends the field by considering the detection of GW events from a \emph{nonstationary} noise background representative of realistic LIGO detectors. In addition to the main GW readout, we further show that in principle one can also input to the NN some key auxiliary channels witnessing the sources of contamination to hence enhance the low-frequency sensitivity. As the contamination typically involves nonlinear and nonstationary coupling mechanisms, it cannot be mitigated by standard signal processing techniques assuming linear and stationary noise coupling. 
We demonstrate that with NNs involving nonlinear activations,  one can nonetheless tackle the challenges of nonlinearity and nonstationarity and achieve simultaneous noise mitigation and signal detection both in real time.  

The rest of the paper is organized as follows. In Sec.~\ref{sec:sensitivity_overview} we briefly overview the LIGO sensitivity during its third observing run (O3) and discuss the main source of contamination to the low frequency band of interest. In Sec.~\ref{sec:gw_overview} we then describe the properties of the GW signal. This is followed by Sec.~\ref{sec:neural_network} in which we provide the details of the construction of training of our early-warning NN. Specifically, we describe the preparation of our training datasets in Sec.~\ref{sec:data_prep} and then in Secs.~\ref{sec:noise_sub}-\ref{sec:combined_prob} the procedures we adopt for the network training. The performance of our NN is examined in Sec.~\ref{sec:results}. Lastly, we conclude and discuss our results in Sec.~\ref{sec:conclusion}

\section{Overview of LIGO sensitivity}
\label{sec:sensitivity_overview}

\begin{figure}
  \centering
  \includegraphics[width=0.9\columnwidth]{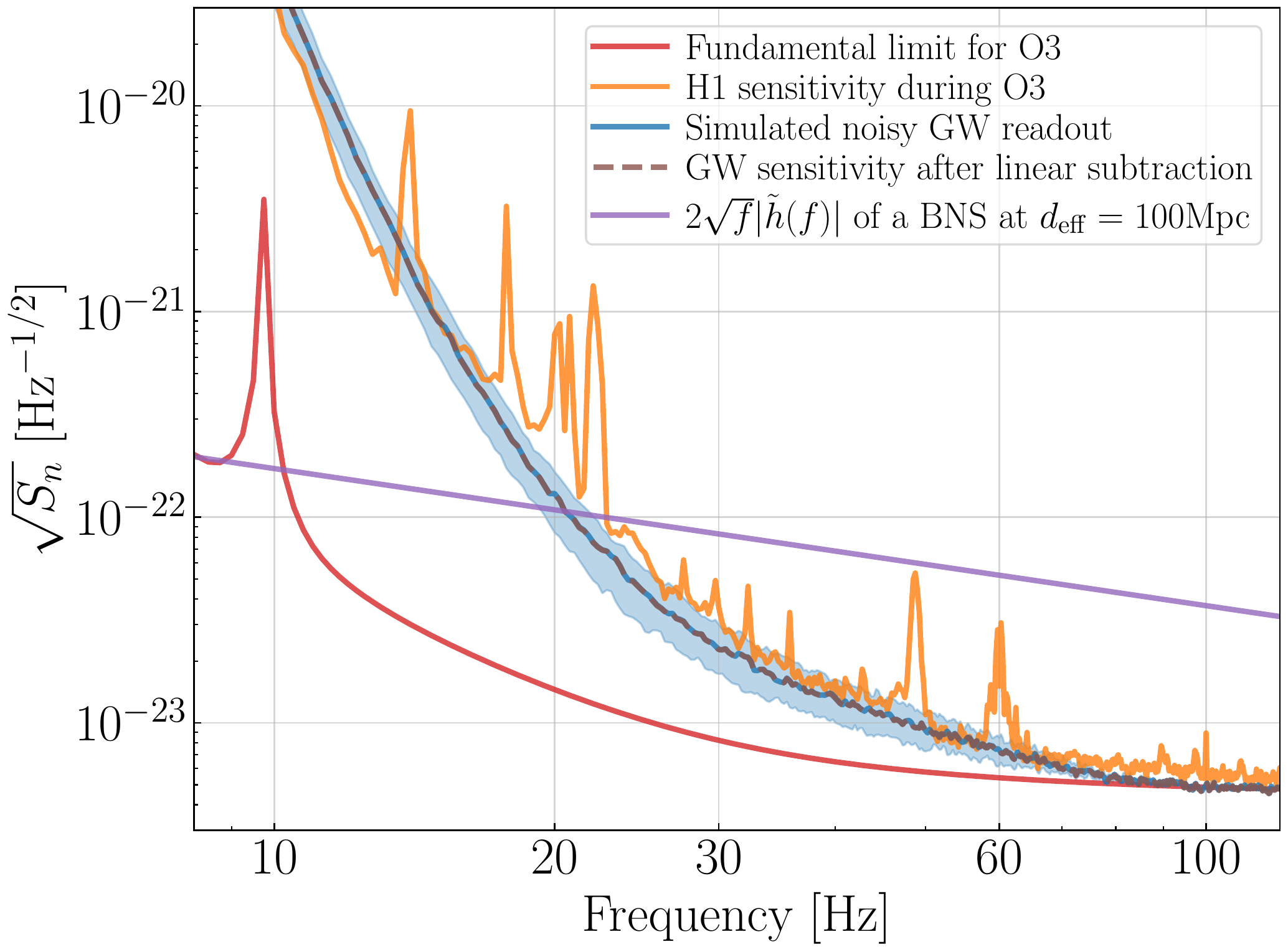}
  \caption{Comparison of the strain sensitivities. The orange trace is the realistic sensitivity of the LIGO Hanford detector during O3~\cite{Buikema:20} and the red trace its fundamental limit. We simulate noise according to the mechanism described in Eq.~(\ref{eq:bilin_noise}) and a typical realization is shown in the blue trace. Note that it is in fact nonstationary and its spectrum can vary within the blue-shaded band. Because of the nonlinear nature of the noise coupling, a linear, coherence-based subtraction cannot mitigate the noise as shown in the dashed-brown trace. As a reference, we also show the strain of a typical coalescing BNS event in the purple trace. The event can be detected at a much lower frequency (hence a much earlier time) if the contamination at low frequencies can be mitigated.}
\label{fig:ASDs_comp}
\end{figure}

While LIGO has achieved a great success, its sensitivity can still be further improved as we demonstrate in Fig.~\ref{fig:ASDs_comp}. Here the orange trace is the representative sensitivity at the LIGO Hanford observatory during the third observing run (O3)~\cite{Buikema:20} and the red trace is its fundamental limit set by quantum and thermal fluctuations at the O3 configuration (which actually closely matches the designed sensitivity of aLIGO~\cite{TheLIGOScientific:2014jea}). While the two traces overlaps at $f\gtrsim 100\,{\rm Hz}$, at lower frequencies the excess noise can be significant. At 30\,Hz (20\,Hz), the fundamental limit is a factor of 3 (10) below the current sensitivity, indicating a large room of improvement. Opening up the sensitivity in the low-frequency band can be especially rewarding for multi-messenger astronomy and astrophysics, as it allows a coalescing BNS (whose strain we show in the purple trace) to be detected at a lower frequency and hence a much earlier time prior to the merger; see the discussion in Sec.~\ref{sec:gw_overview}. 

A major source of contamination to the current low-frequency sensitivity is the control noises of auxiliary degrees of freedom~\cite{Buikema:20} (see also, e.g.,  Refs.~\cite{Martynov:16, Yu:18}). 
For instance, while it is necessary to engage an active angular control system to maintain the alignment of test masses at below a few Hz during LIGO's observation, the system also inevitably feeds back the sensing noise in the 10-30\,Hz band and causes excess angular perturbation $\theta(t)$. The angular perturbation further couples with the off-pivot beam spot motion and leads to a longitudinal displacement that contaminates the GW readout as 
\begin{equation}
    \delta x(t) = x_{\rm spot} (t) \theta (t) = \left[x_{\rm spot}^{\rm (DC)} + x_{\rm spot}^{\rm (AC)} (t)\right] \theta (t).
    \label{eq:bilin_noise}
\end{equation}
Such a contamination can be mitigated by both online feed-forward cancellation and offline signal regressions (see, e.g., Ref.~\cite{Driggers:19}). However, standard signal processing techniques [such as computing the Wiener filter from $\theta(t)$ to $\delta x(t)$] assume the coupling is \emph{linear and stationary} and therefore can only remove the constant coupling part $\propto x_{\rm spot}^{\rm (DC)}$ but not the fluctuating piece $\propto x_{\rm spot}^{\rm (AC)} (t)$. In fact, it is exactly due to the temporal variability of couplings like $x_{\rm spot}^{\rm (AC)} (t)$  that the current LIGO noise background at low frequencies is \emph{nonstationary}~\cite{Martynov:16, Buikema:20}, dramatically complicating the data analysis process~\cite{Zackay:2019kkv}. 

Furthermore, there are no direct witnesses for the spot position on the test masses, $x_{\rm spot}^{\rm (AC)}(t)$, over the entire frequency band of interests. Instead, it has to be reconstructed from multiple sensors through complicated geometrical conversions as well as signal filtering and blending, with each step subject to its own calibration uncertainties.

Nonetheless, neural networks (NNs) using machine learning (ML) offers an attractive way to tackle this problem. By inputting sufficient auxiliary witness channels, a deep convolutional neural network (CNN)~\cite{Lecun:98} would be able to figure out the correct, frequency-dependent combinations of the witness that reconstructs the contamination. Moreover, as each layer typically involves a nonlinear activation function, it would be able to capture \emph{nonlinear} couplings like Eq.~(\ref{eq:bilin_noise}) that classical, \emph{linear} signal processing techniques fail (see also Refs.~\cite{DaSilvaCosta:18, Mukund:20, Vajente:20, Ormiston:20} for some recent efforts to mitigate nonlinear noises in the LIGO detectors). Furthermore, as an NN is trained directly on time series, it is especially suitable to be implemented in real-time and has the potential to be integrated into a low-latency detection pipeline. 

To demonstrate this point, we simulate excess noise according to the mechanism described in Eq.~(\ref{eq:bilin_noise}) and combine it with the fundamental limit to form the blue trace in Fig.~\ref{fig:ASDs_comp}. The $x_{\rm spot} (t)$ and $\theta(t)$ as well their witness channels are simulated with similar characteristics as in realistic LIGO detectors, with one exception that we reduce the roll-off of $\theta(t)$ in the 25-80\,Hz band so that the entire O3 sensitivity can be approximated by this mechanism (see Sec~\ref{sec:data_prep} for more details). In reality, the noise in the 25-80\,Hz are dominated by other noise sources~\cite{Buikema:20} which we ignore here for simplicity. Note that we have assumed that the constant coupling piece is already removed (i.e., $x_{\rm spot}^{\rm (DC)}=0$), and linear subtraction cannot further mitigate the contamination. This is illustrated by the brown-dashed curve in Fig.~\ref{fig:ASDs_comp} where we compute the multi-input-single-output coherence between all the auxiliary witness and the gain GW channel and then subtract out the coherent component in the frequency domain. To further simulate the nonstationarity on timescales longer than the length of each realization (256\,s), we allow the overall root-mean-square (RMS) of $x_{\rm spot} (t)$ to be a random variable. Thus the blue trace in Fig.~\ref{fig:ASDs_comp} is just the ASD of a typical realization; the noises we simulate in fact has their spectra vary within the shaded blue region (see also Fig.~\ref{fig:noise_residual}).

\section{GW signal}
\label{sec:gw_overview}
Having described the noise and how we may use ML techniques to mitigate it, we now turn to the discussion about detecting the astrophysical GW events. Specifically, our goal is to detect a GW event minutes before the final merger and further classify its type (NS vs BH) to assist the EM follow up strategies. 

For the early warning purpose, we can approximate the waveform using only the leading-order quadrupole formula and write (with $G=c=1$; see, e.g., \cite{Maggiore:08}) 
\begin{align}
    &h(t) = \frac{\mathcal{A}}{d} \mathcal{M}_c^{5/4} \left(\frac{5}{t_m}\right)^{1/4} \cos[\Phi(t)], \label{eq:hoft}\\
    &\Phi(t) = -2 \left(\frac{t_m}{5\mathcal{M}_c}\right)^{5/8} + \Phi_{\rm c},
\end{align}
where $t_m = t_c - t$ is the time to merger and  $t_c$ and $\Phi_c$ are time and phase of coalescence. The time $t_m$ is further related to the GW frequency $f$ according to
\begin{align}
    &t_m(f) = 86.7\,{\rm s}\left(\frac{\mathcal{M}_c}{1.22 M_\odot}\right)^{-5/3}\left(\frac{f}{25\,{\rm Hz}}\right)^{-8/3}, \label{eq:tm_vs_f}\\
    &f(t_m) = 23.7\,{\rm Hz}\left(\frac{\mathcal{M}_c}{1.22 M_\odot}\right)^{-5/8} \left(\frac{t_m}{100\,{\rm s}}\right)^{-3/8}.\label{eq:f_vs_tm}
\end{align}
In this work, we do not include the detailed antenna responses (which are encoded in the quantity $\mathcal{A}$) nor the joint detection by multiple detectors. Instead,  we set $\mathcal{A}=1$ and simply replace the distance to the source $d$ in Eq.~(\ref{eq:hoft}) by $d_{\rm eff}/\sqrt{N_{\rm det}}$, where $d_{\rm eff}\simeq 2.3d$ is the averaged effective distance~\cite{Finn:93, Allen:12} and $N_{\rm det}$ is the number of detectors observing. For the rest of the work, we will use $N_{\rm det}=3$ as the default value.

From the above equations we see that the waveform depends only on one intrinsic parameter of the source, the chirp mass $\mathcal{M}_c$, defined as 
\begin{equation}
    \mathcal{M}_c = \frac{(M_1 M_2)^{3/5}}{(M_1 + M_2)^{1/5}},
\end{equation}
with $M_{1,2}$ the component masses. Therefore, we put a GW event into three categories according to its chirp mass. 

We define the first category as events with $1\,M_\odot\leq \mathcal{M}<1.8$ and label such an event as a ``BNS'' event. Note that a BNS with $M_1=M_2=2\,{M_\odot}$ (which is the mass of the heaviest NS observed to date~\cite{Cromartie:20}) will have $\mathcal{M}_c=1.74\,M_\odot$. Therefore, we would expect that most astrophysical BNS events will fall into this category (including GW170817with $\mathcal{M}_c=1.19\,M_\odot$~\cite{GW170817a} and GW190425 with $\mathcal{M}_c=1.44\,M_\odot$~\cite{GW190425}). 

\begin{figure}
  \centering
  \includegraphics[width=0.9\columnwidth]{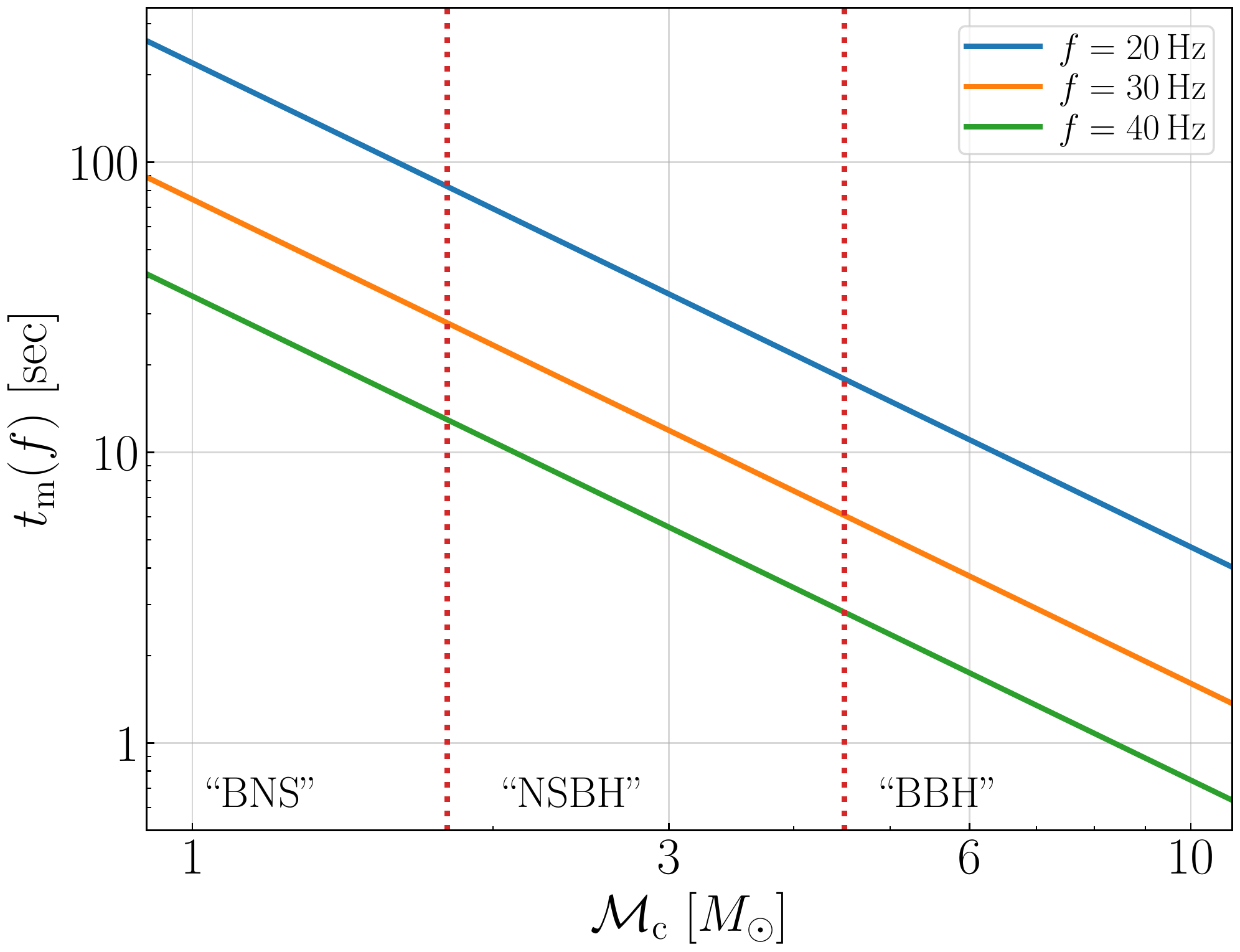}
  \caption{Merger time as a function of the chirp mass. The vertical lines correspond the boundaries of the three signal classes we consider here. Note that the label for each class is put under quotation marks because here we only loosely define each class by its chirp mass, the information that is best constrained at the early inspiral stage. The source's properties can be better refined by follow-up analysis utilizing data at higher GW frequencies. }
\label{fig:tm_vs_Mc}
\end{figure}

We also define a ``BBH'' category as sources with $4.5\,M_\odot\leq \mathcal{M}_c < 10\,M_\odot$. The lower boundary is inspired by noticing a BBH with $M_1=M_2=5\,M_\odot$ would have $\mathcal{M}_c=4.35\,M_\odot$. In principle, the upper boundary of $\mathcal{M}_c < 10\,M_\odot$ for this category is not necessary (or it should be set to a much greater value). We nonetheless put it to $10\,M_\odot$ for the training simplicity. Moreover, more massive systems merges in only a few seconds or even less in duration [Eq.~(\ref{eq:tm_vs_f}) and Fig.~\ref{fig:tm_vs_Mc}], and therefore they are not the main target of our study here.

Lastly, we refer to sources with $1.8\,M_\odot\leq \mathcal{M}_c < 4.5\,M_\odot$ as the ``NSBH'' category, as it covers events with $(M_1, M_2)=(8\,M_\odot, 1.4\,M_\odot)$, or $\mathcal{M}_c=2.7\,M_\odot$. We nonetheless point out that this category may also contain a binary of NSs both massive than $2\,M_\odot$ or a pair of light BHs both in the lower ``mass gap'' with $M_{1,2}<5\,M_\odot$. While it is possible to refine our knowledge of the source if we include dynamics at high post-Newtonian orders and/or potential tidal interactions, these effects are encoded at higher GW frequencies and therefore is beyond the scope of our work targeting the early warning using only the low-frequency portion of the signal. Indeed, at the frequency range we are interested in here, the corrections we drop is only on the order $\sim v^2\simeq 1.1\%\left[(M_1+M_2)/3\,M_\odot\right]^{2/3}(f/25\,{\rm Hz})^{2/3}$.  Nonetheless, we may imagine our work here would serve as a first step of a future, integrated early warning pipeline, and once an event is detected here, it can then trigger further analysis on the signal to refine its property.

In Fig.~\ref{fig:tm_vs_Mc} we show the merger time $t_m$ for binaries with different chirp masses $\mathcal{M}_c$ at three different GW frequencies. The two vertical, dotted lines indicate the boundaries between the three categories defined in our study. From the plot we see that if the event can be detected by $\simeq 30\,{\rm Hz}$, then for the three categories (``BNS'', ``NSBH'', ``BBH''), we should in principle be able to detect the signal $\left[\mathcal{O}(100), \mathcal{O}(10),\mathcal{O}(1)\right] \,{\rm s}$ prior to the merger. 

\begin{figure}
  \centering
  \includegraphics[width=0.9\columnwidth]{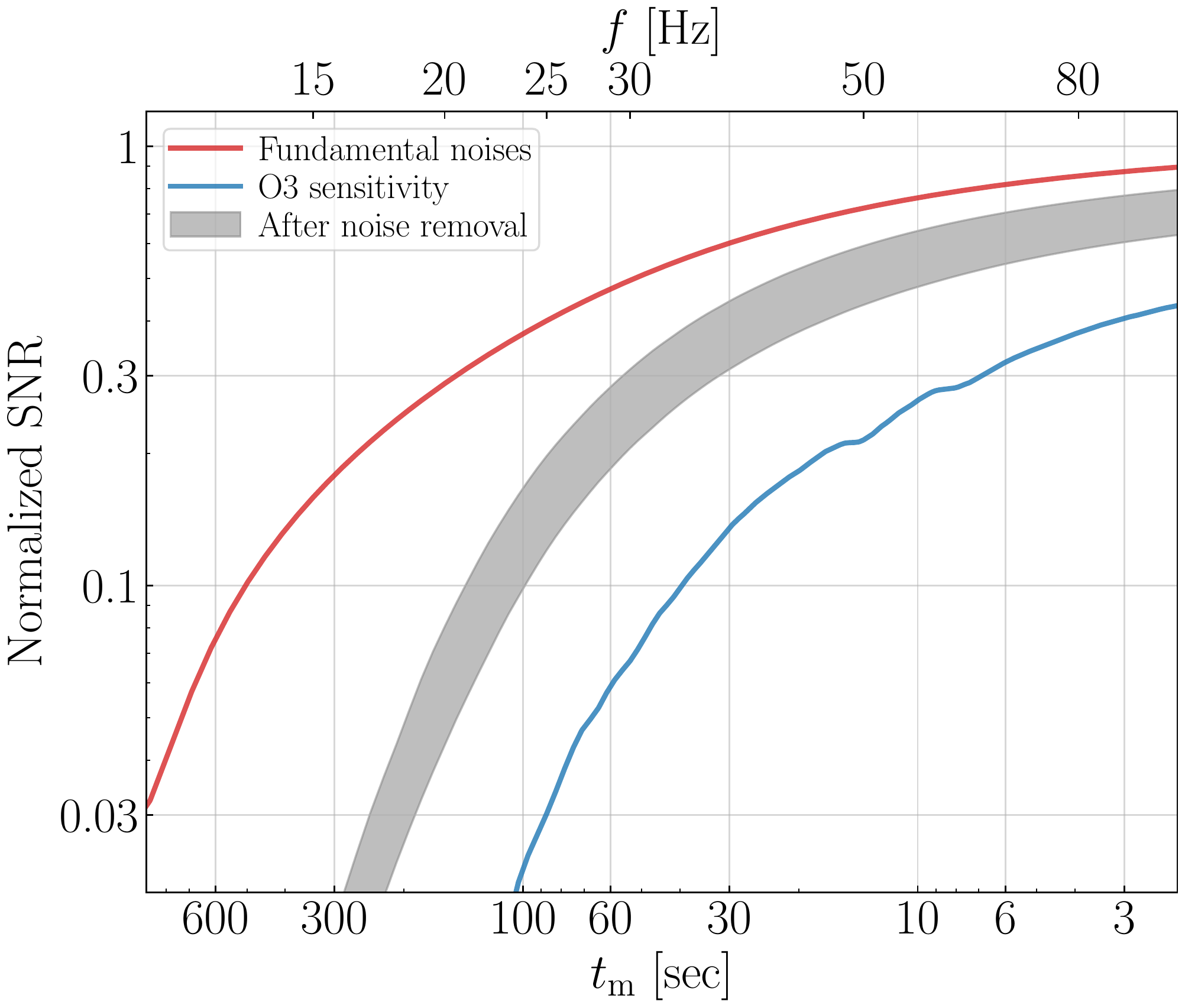}
  \caption{Cumulative SNR as a function of $t_m$ (bottom x-axis) and $f$ (top x-axis) for a BNS with $\mathcal{M}_c=1.22\,M_\odot$. The red trace is computed using the fundamental sensitivity and the blue one using the realistic O3 sensitivity. Also shown in the grey band is the SNR using the residual noises after the CNN's cleaning (Sec.~\ref{sec:noise_sub}). At 100\,s prior to the merger, one could in principle integrate about 40\% of the total SNR if we reach the design sensitivity, yet currently only about 1\% is accumulated in this band.}
\label{fig:BNS_cumSNR}
\end{figure}

In reality, the situation may be more challenging because the current LIGO low-frequency sensitivity is orders of magnitude above its fundamental limit as we have already seen in Fig.~\ref{fig:ASDs_comp}. We illustrate this point further in  Fig.~\ref{fig:BNS_cumSNR} where we show the cumulative signal-to-noise ratio (SNR) $\rho$ for a BNS with $M_1=M_2=1.4\,M_\odot$ as a function of $t_m$ (bottom x-axis) and $f$ (top x-axis). Specifically, we define the cumulative SNR through
\begin{align}
    \rho^2 [f<f(t_m)]
    {{=}} &4{\rm Re}\left[\int^{f(t_m)} \frac{\tilde{h}^\ast(f)\tilde{h}(f)}{S_n(f)} df\right],
    \label{eq:snr}
\end{align}
where $\tilde{h}(f)=\int h(t) \exp\left(i2\pi f t\right) dt$ and $S_n$ the detector's power spectral density. In the plot, we further normalize the curves by the total SNR assuming the fundamental O3 sensitivity (the red trace in Fig.~\ref{fig:ASDs_comp}). 

As can be seen from Fig.~\ref{fig:BNS_cumSNR}, with the current O3 sensitivity (blue trace), to accumulate to a normalized SNR of 0.2, we have to integrate the signal to around 40 Hz or $t_m\simeq 20\,{\rm s}$. Such a time window might not be sufficient especially if one wants to catch potential precursor signals of BNS mergers given various realistic delays in the information communication and decision making. In contrast, if LIGO can reach its fundamental limit, one would only need to integrate to 15\,Hz, which is 300\,s prior to the merger. It thus demonstrates the great scientific reward of enhancing the low-frequency sensitivity, which we propose to achieve via ML-based nonlinear noise regression.

\section{Neural network}
\label{sec:neural_network}
\begin{figure}
  \centering
  \includegraphics[width=0.95\columnwidth]{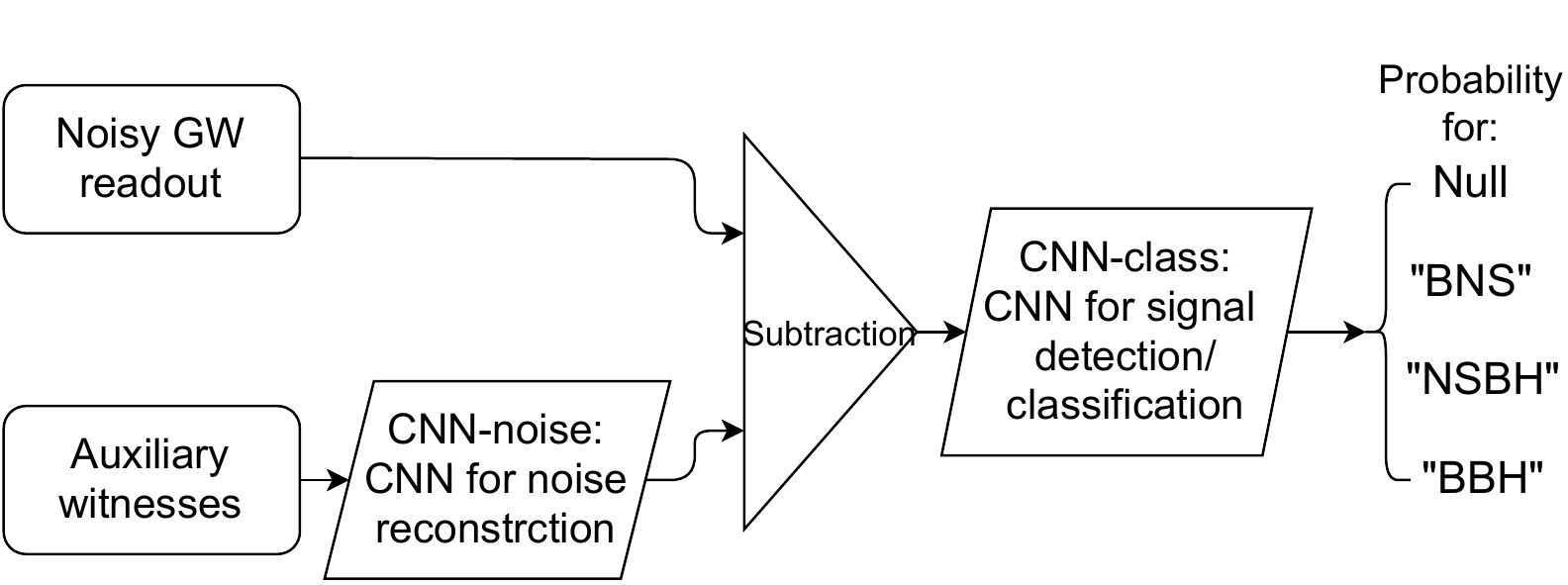}
  \caption{A compound CNN we propose to use to detect GW events. In the ``CNN-noise'' CNN model we first reconstruct the noise that limits the low-frequency sensitivity using auxiliary channels. We then subtract its output from the main GW readout and then pass the residuals to ``CNN-class'' model which outputs the probability of the input time-series belonging to each one of the classes we defined. The two CNNs can be first trained individually and then combined and optimized globally.}
\label{fig:compound_CNN}
\end{figure}

A cartoon illustrating the proposed NN structure is shown in  Fig.~\ref{fig:compound_CNN}. Here we input both the main GW readout and a few key auxiliary witness channels to simultaneously achieve noise mitigation and signal classification.

To assist the convergence of the network, we adopt a compound structure. We first use the network ``CNN-noise'' to preform noise reconstruction and then subtract its output from the noisy GW readout to form a cleaned strain signal. This is then fed to the network ``CNN-class'' to achieve signal detection and classification. Both sub-networks can be first trained individually and then combined together to preform a global optimization. 

Our ML training is preformed using \texttt{Keras}~\cite{Chollet:15}, a \texttt{Python}-based interface running on top of the ML platform \texttt{TensorFlow}~\cite{Abadi:15}. The details of data generation and network training is presented below. 

\subsection{Data preparation}
\label{sec:data_prep}

We describe in this Section how we generate the data we used for training the CNN.

We generate the GW signal from Eq.~(\ref{eq:hoft}) with the distance $d$ replaced by $2.3d_{\rm eff}/\sqrt{N_{\rm det}}$ and $N_{\rm det}=3$. For training the NN, it is not necessary to sample the masses following a specific astrophysical distribution. 
Instead, we sample $M_c$ from a normal distribution with a mean of $1.22\,M_\odot$ and standard deviation of $0.3\,M_\odot$ and truncate the distribution at $[1, 1.8)\,M_\odot$, the predefined range of the ``BNS'' class. For ``NSBH'' and ``BBH'', the masses are simply sampled from uniform distributions. 

To achieve early warning, we do not use the entire waveform up to the merger, but truncate the high-frequency end of the waveform at a cutoff frequency $f_{\rm cut}$. For the ``BNS'' class, we randomly sample $f_{\rm cut}$ between 24\,Hz and 25\,Hz. For typical BNS event with $M_1=M_2=1.4\,M_\odot$ ($\mathcal{M}_c=1.2\,M_\odot$), this corresponds to $t_{\rm m}(f_{\rm cut})=97-87\,{\rm s}$. The starting frequency is chosen such that the integration time $t_{\rm int}$ of the signal is 256\,s. Because more massive systems can evolve to higher frequencies for a given amount of integration time, we set $f_{\rm cut}$ to slightly higher values for the ``NSBH'' and the ``BBH'' classes; we sample $f_{\rm cut}$ from $[28,32)\,{\rm Hz}$ and $[35, 40)\,{\rm Hz}$. For an NSBH event with $(M_1, M_2)=(8\,M_\odot, 1.4\,M_\odot)$ this leads to 17\,s to 12\,s of pre-merger warning time, and for a BBH with $M_1=M_2=5\,M_\odot$, it is 4\,s to 3\,s prior to the merger. The integration time $t_{\rm int}$ is  the minimum of 256\,s  and $t_{m}(10\,{\rm Hz}) - t_{m}(f_{\rm cut})$. In all cases, the phase at coalescence $\Phi_c$ is always sampled randomly from $[0, 2\pi)$. Because we consider $f_{\rm cut}<40\,{\rm Hz}$, we only need to sample each waveform at a rate of 256\,Hz. Such a relatively low sampling rate is the key allowing us to integrate the signal for a duration as long as 256\,s.
In Table~\ref{tab:sig_class_def} we summarize the key parameters of the three signal classes we consider.

Once a waveform is generated, we then inject it to a noise background of 256\,s long, containing both a stationary part due to the fundamental noise limit (the red trace in Fig.~\ref{fig:ASDs_comp}), and an additional low-frequency contamination represented by the blue stripe in Fig.~\ref{fig:ASDs_comp} (see the description shortly after). As one may imagine continuously passing the strain data to the CNN we trained here (specifically, ``CNN-class'' in Fig.~\ref{fig:compound_CNN}), we align the signal so that it reaches $f_{\rm cut}$ at the end of the time series. 

Together with the three signal classes, we also consider a ``null'' class containing only the nonstatinary detector noise. The goal of CNN-class is then to output the probability of a 256-second data series belonging to one of the four classes. 

\begin{table}
\caption{Summary of the three signal categories considered in our study. The first row is the chirp mass defining each category. The second row is the frequency at which we cut off the signal; only signal prior to this cutoff is used for detection. The third row is the typical time to merger at the cutoff frequency and the last row is the integration time of each signal. 
The chirp masses are given in $[M_\odot]$, times are in $[{\rm sec}]$, and frequencies are in $[{\rm Hz}]$. All waveforms are sampled at a rate of 256\,Hz. \label{tab:sig_class_def}}
\begin{ruledtabular}
\begin{tabular}{lccc}
\textrm{label}&
\textrm{``BNS''}&\textrm{``NSBH''}&\textrm{``BBH''}
\\
\colrule
$\mathcal{M}_c$ & [1, 1.8) & [1.8, 4.5) & [4.5, 10)\\
$f_{\rm cut}$ & [24, 25) & [28, 32) & [35, 40)\\
$t_{\rm m}^{\rm typ}|_{f_{\rm cut}}$ & [97, 87) & [17, 12) & [4,3)\\
$t_{\rm int}$ & 256 & ${\rm min}[256, t_m|_{10}{-}t_m|_{f_{\rm cut}}] $ & $t_m|_{10}{-}t_m|_{f_{\rm cut}}$\\
\end{tabular}
\end{ruledtabular}
\end{table}

To simulate the excess low-frequency contamination [blue stripe in Fig.~\ref{fig:ASDs_comp} and Eq.~(\ref{eq:bilin_noise})], we generate noises with similar characteristics as in realistic LIGO detectors. 

Specifically, we simulate four independent time series of the fast ($\gtrsim 10\,{\rm Hz}$) angular motion $\theta(t)$, corresponding to sensing noises in the four high-bandwidth angular feedback loops (for controlling pitch and yaw motions of the two arm cavities).\footnote{There are four instead of eight independent degrees of $\theta(t)$ because only a specific linear combination of the input and end test masses, known as the ``hard mode''~\cite{Sidles:06}, requires a high control bandwidth ($\sim 3\,{\rm Hz}$) and thus injects a significant amount of sensing noise at $\gtrsim 10\,{\rm Hz}$~\cite{Buikema:20}. The orthogonal combination, or the ``soft mode'', only requires to be  controlled with a bandwidth of $\sim 0.5\,{\rm Hz}$ and has negligible contribution to the noise. } 
Instead of using realistic spectral shapes for $\theta(t)$ as in the LIGO system, we design them so that the contamination has a spectral shape similar to the full O3 sensitivity~\cite{Buikema:20}. In other words, we give $\theta(t)$ extra power in the $>25\,{\rm Hz}$ band and ignore other sources of contamination in this band. This does not affect the main results of our study though because we choose $f_{\rm cut}$ between 24\,Hz and 25\,Hz for the ``BNS'' class. 

Meanwhile, we also simulate eight independent spot-position motions $x_{\rm spot}(t)$ for the four test masses and two angular degrees of freedom (pitch and yaw). Their motions are mostly induced by the microseismic motion and peak in the $0.1-0.3\,{\rm Hz}$ band. This is the main source of nonstationarity on timescale of $10\,{\rm s}$. 
At longer timescales, the overall RMS value of $x(t)$ drifts and shows seasonal dependence: during winter times the microseimic motion is typically higher than in the summer. 
To simulate this, on top of a typical value ${\rm RMS}\left[x(t)\right]\simeq 0.3\,{\rm mm}$,  we additionally sample an overall scale factor uniformly from $[0.7, 1.4]$ and apply it to $x(t)$ for each realization.

In order to sense the true spot motions, we assume the information is contained in two sets of witness sensors. The first set of sensors probe the spot motion by exciting each mirror in angle and looking for length fluctuations at the excitation frequency. The angle-to-length conversion factor directly gives us the spot motion at each test mass [see Eq.~(\ref{eq:bilin_noise})]. However, they have very limited SNR and can only trace the long-term ($\lesssim0.1\,{\rm Hz}$) drift of the spot motion. The other set of sensors are optical levers placed locally at each test mass. They senses the angular motion of each test mass relative to its local ground, which can then be converted to the spot motion using the cavity's geometry. They provide information in the $\gtrsim 0.1\,{\rm Hz}$ band but are polluted by seismic and thermal drifts at lower frequencies and therefore are not coherent with the true spot motion at $< 0.1\,{\rm  Hz}$. Consequently, we would need two sensors (one dithering-based sensor and one optical lever) for the spot motion per test mass per direction. In total, we thus need 20 auxiliary witness channels [4 for $\theta(t)$ and 16 for $x_{\rm spot}(t)$] to reconstruct the low-frequency contamination
\footnote{More specifically, we simulate the fast channels ($\theta(t)$) with similar characteristics as LIGO auxiliary channels like \texttt{H1:ASC-CHARD\_P\_OUT\_DQ}. The slow channels are designed to mimic auxiliary channels like \texttt{H1:ASC-ADS\_PIT4\_DOF\_OUT\_DQ} for the dithering based spot position sensors, and \texttt{H1:SUS-ITMX\_L3\_OPLEV\_PIT\_OUT\_DQ} for the optical lever outputs.
}. 
Same as the main GW readout, all of the auxiliary channels are sampled at a rate of 256 Hz.

To reduce the complexity of the problem, we first train the two sub models, ``CNN-noise'' and ``CNN-class'', individually, which we will describe in Sec.~\ref{sec:noise_sub} and Sec.~\ref{sec:sig_class}. After each sub-model's convergence,  we then load their weights  into the compound model as the initial condition and preform a global optimization (Sec.~\ref{sec:combined_prob}). 

\subsection{Noise subtraction}
\label{sec:noise_sub}

\begin{table}
\caption{Network structure for noise subtraction. The network includes about 640,000 trainable parameters in total.
\label{tab:CNN_noise}}
\begin{ruledtabular}
\begin{tabular}{cccc}
\textrm{layer}&
\textrm{output dimension}&\textrm{kernel size}&\textrm{activation}
\\
\colrule
Conv1D & 256 & 64 & ELU \\
Conv1D & 32 & 16 & ELU \\
Dropout & -- & -- & -- \\
Conv1D & 256 & 8 & ELU \\
BatchNormalization & -- & -- & -- \\
Dense & 256 & -- & ELU \\
Dropout & -- & -- & -- \\
Dense & 128 & -- & ELU \\
Dense & 128 & -- & ELU \\
Dense & 16 & -- & ELU \\
BatchNormalization & -- & -- & -- \\
Dense & 1 & -- & Linear \\
\end{tabular}
\end{ruledtabular}
\end{table}

\begin{figure}
  \centering
  \includegraphics[width=0.9\columnwidth]{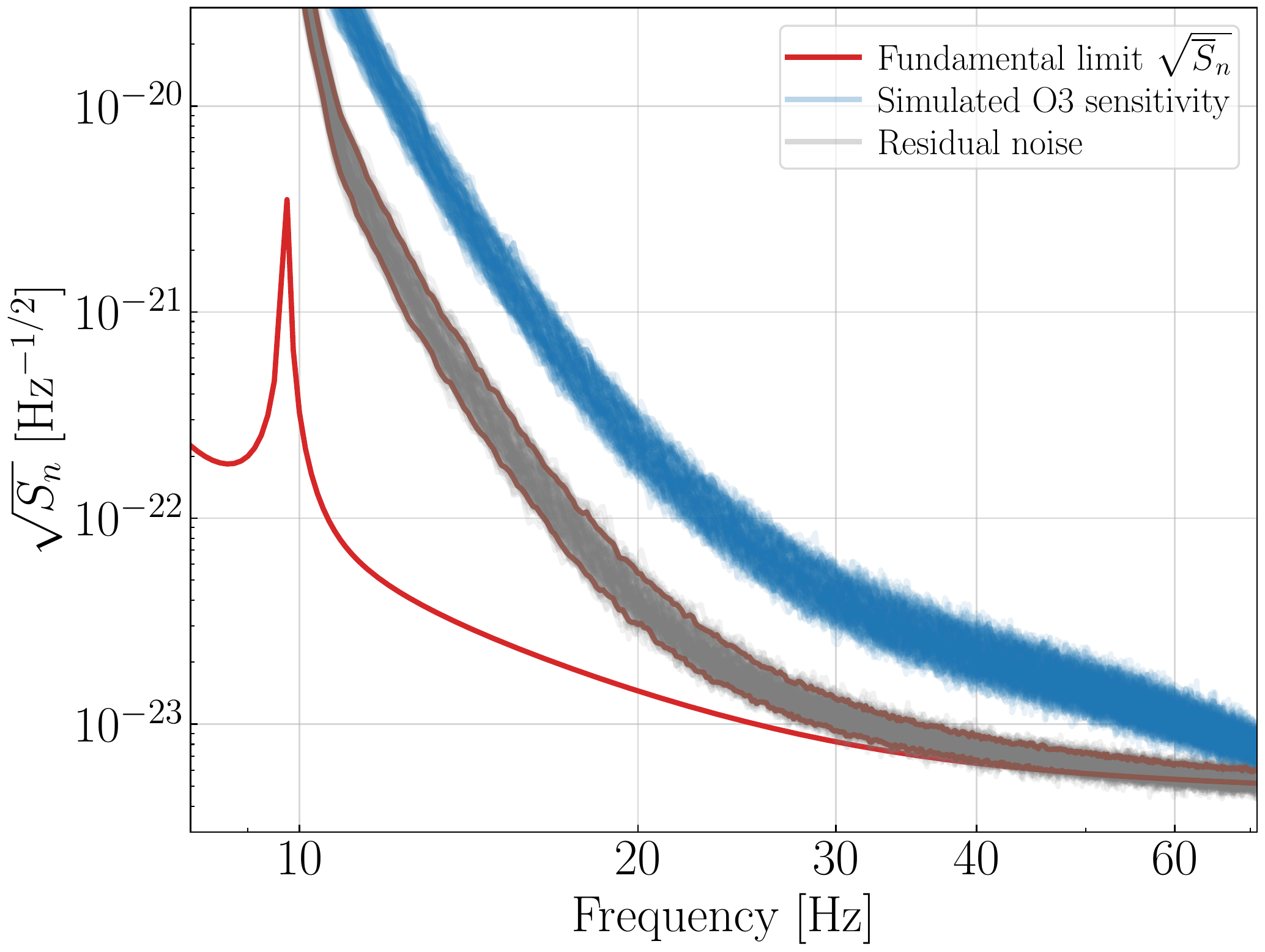}
  \caption{Residual noise after the cleaning by CNN-noise. Each blue trace corresponds to one realization of our simulated detector noise and the grey ones the residual after noise subtraction. The two brown traces correspond to the 5- and 95-percentiles of the residual. With the current network, we can achieve a factor of 5 broadband subtraction of the noise.}
\label{fig:noise_residual}
\end{figure}

Our first step is to construct a NN that mitigates the excess low-frequency contamination to the GW readout in real time. We will refer to this NN specifically as ``CNN-noise''.  It takes the 20 auxiliary witness channels we simulate as the input and estimates their nonlinear contamination to the main GW readout as the output (see also Fig.~\ref{fig:compound_CNN}). To achieve supervised learning, we use time series from the noisy GW readout as training targets for this step. 

Because for most of the observation time there will be no GW signal present in the data, the training is thus preformed on signal-free data series only in this step. We also do not need to use the full 256 seconds of data for each training segment for noise mitigation because the contamination relies only on the instantaneous spot and angle [Eq.~(\ref{eq:bilin_noise})]. The time series only needs to be long enough to capture the microseismic motion (with a characteristic period of $\sim 10\,{\rm s}$) which is the main cause of fluctuations in the spot motion. Consequently, we use 64 seconds of data from 21 channels (20 auxiliary witnesses as the input and 1 noisy GW readout as the target) for each segment (i.e., ``batch'' in the ML literatures), and train ``CNN-noise'' over 128 segments for each training epoch. 

Moreover, for the convenience of the subsequent signal classification, we would like the ``cleaned'' GW readout to have a nearly white spectral shape. Therefore, we precondition the noisy GW readout before it is passed for training. Since the current O3 detector noise is orders of magnitude greater than the fundamental noise limit (the ideal output of noise cleaning) at $f\lesssim 20\,{\rm Hz}$, the precondition is done in an iterative way. 

In the first iteration, we whiten the GW readout according to the fundamental O3 noise limit. The spectrum of the residual after noise subtraction is then used to design the preconditioning filter for whitening the GW readout in the next iteration. Because of the nonlinearity involved in the noise coupling, a CNN trained to estimate $\delta x(t)$ from $\left[\theta(t), x_{\rm spot}(t)\right]$ according to Eq.~(\ref{eq:bilin_noise}) does not apply for approximating $\mathcal{L}\left\{\delta x(t)\right\}$ from  $\left[\mathcal{L}\left\{\theta(t)\right\}, x_{\rm spot}(t)\right]$ with $\mathcal{L}$ denoting a generic linear operator. Therefore, the weights in ``CNN-noise'' need to be updated once the preconditioning filter changes. Nevertheless, we find the residual are similar for the first and second iterations, and therefore we do not to iterate further. 

The same preconditioning filter is also applied to the witness channels for $\theta(t)$. While this does not preserve the exact coupling as we argued above, we nonetheless find it helps the CNN to converge faster numerically. 

As for the witness channels for the spot motion, we only apply an overall calibration factor so that each channel's numerical values are of order unity. Specifically, we calibrate the dithering-based sensors to output spot motion in millimeter and the optical levers to output the low-frequency ($<1\,{\rm Hz}$) angular motion \footnote{Note the angular motion sensed by the optical levers are due to low-frequency seismic motion. It is the source for spot-position motion $x_{\rm spot}(t)$ and should be distinguished from the high-frequency ($>10\,{\rm Hz}$) angular motion $\theta(t)$ due to sensing noises in the control feedback.} in microradians. Note that the overall RMS of each channel contains physical meaning [the coupling strength from $\theta (t) $ to $\delta x(t)$] and should not be normalized out. Similarly, for each GW readout we apply a fixed normalization constant. 

Once the data are generated and preconditioned, we then pass them to ``CNN-noise'' to learn the nonlinear noise coupling from the auxiliary channels to the main GW readout. The best performing network structure is summarized in Table~\ref{tab:CNN_noise}. 

We construct a custom loss function for the training. Specifically, we compute the loss as 
\begin{equation}
    {\rm Loss} = \int_{f_{\rm low}}^{f_{\rm high}} w S_n^{\rm (resi)} df,
    \label{eq:cumtom_loss}
\end{equation}
where $S_n^{\rm (resi)}$ is the power spectral density of the residual (i.e., $\textrm{target} - \textrm{prediction}$), and $w$ is a weighting function defined as
\begin{equation}
    w = \mathcal{C} \frac{f^{\alpha}}{S_n^{\rm (trgt)}},
\end{equation}
where $S_n^{\rm (trgt)}$ is the power spectral density of the target and $\mathcal{C}$ is an overall constant so that the initial loss is of order unity. Empirically, we set $(f_{\rm low}, f_{\rm high})=(7.5\,{\rm Hz}, 75\,{\rm Hz})$ and $\alpha=-0.5$. In addition, we also sum a small contribution of the standard mean squared error (about 0.1 to the total loss) to  the custom loss defined in Eq.~\ref{eq:cumtom_loss} to avoid artificial offsets at DC due to numerical over-fitting. 

We note that the loss function defined above aims to achieve a broad-band noise mitigation so that the results of ``CNN-noise'' can be applied for various purpose (signal detection, sky localization, etc.). The optimization for the specific purpose of this work (detecting and classifying BNS $\sim 100\,{\rm s}$ prior to the merger) is left for the final step where we combine CNN-noise and CNN-class to preform global training. 


The resultant ASDs of CNN-noise are shown in Fig.~\ref{fig:noise_residual}. In the figure, each blue trace is the amplitude spectral density of a realization of the simulated O3 sensitivity. Similar to the real detector noise, it has a nonstationary nature as the RMS of $x_{\rm spot}^{\rm(AC)}$ various with time (and different from realization to realization). The residual after removing the contamination predicted by CNN-noise using the 20 auxiliary channels is shown in the grey trace. Overall, the contamination can be mitigated by a factor of $\sim 10$, which is sufficient to reach the fundamental limit in the $>30\,{\rm Hz}$ band. At lower frequencies, $f<20\,{\rm Hz}$, even the residual is still an order of magnitude or more above the fundamental limit and the it fluctuates as the spot motion RMS varies, indicating rooms for further improvements. 

\begin{figure}
  \centering
  \includegraphics[width=0.95\columnwidth]{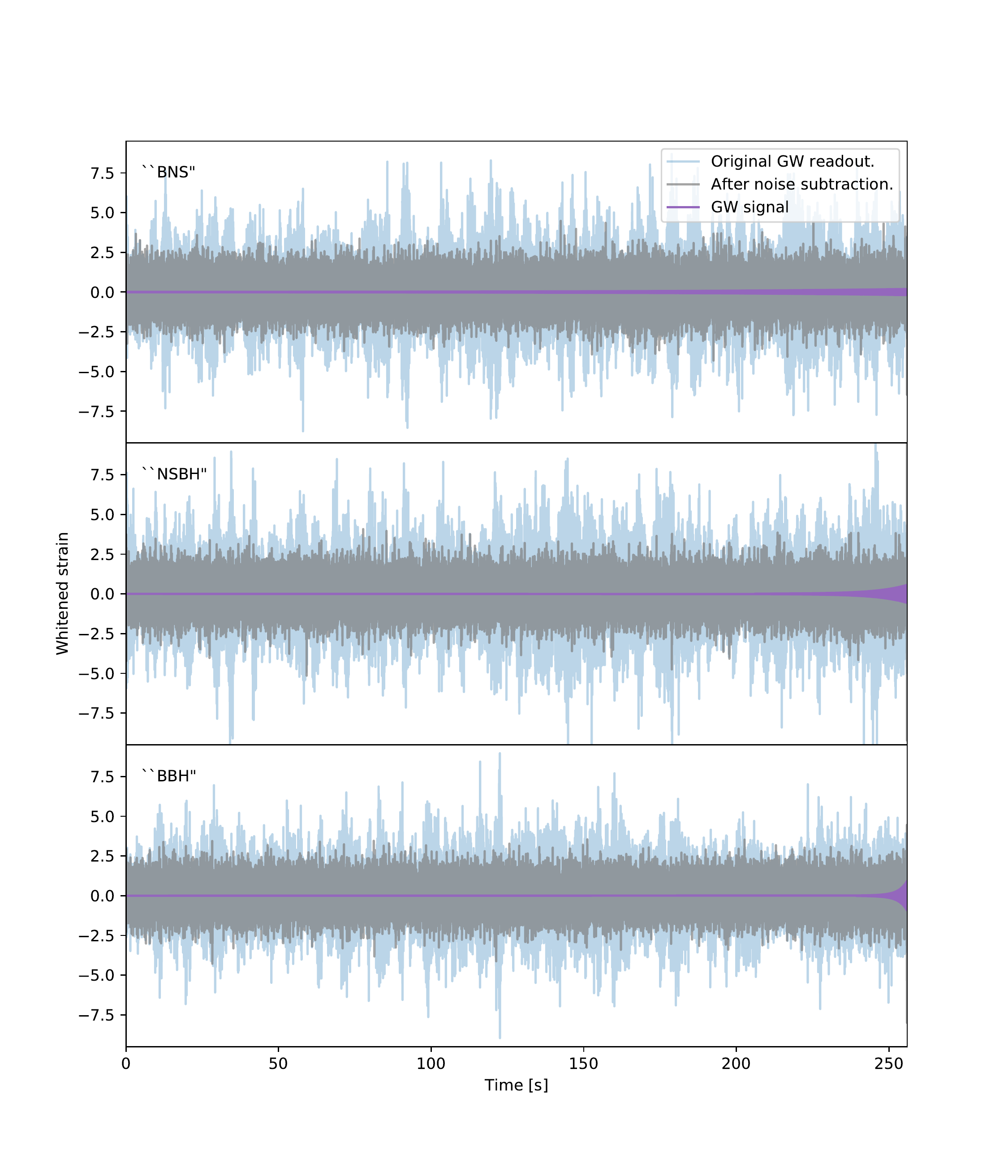}
  \caption{Sample of the whitened time series. The blue traces are the original GW readout simulated according to O3 sensitivity whose band-limited RMS in the [20, 60]\,Hz band varies on a timescale of $10\,{\rm s}$ due to modulations caused by the microseismic motion. The grey traces are the GW readout after noise mitigation by CNN-noise and they are the inputs to CNN-class. The whitened GW signal contained in each realization is highlighted in the purple trace. From top to bottom, they correspond respectively to a typical ``BNS'', ``NSBH'', and ``BBH''. In all the cases we set  $\tilde{\rho}(f<f_{\rm cut})=16$. }
\label{fig:sample_timeseries}
\end{figure}

Note that each curve in Fig.~\ref{fig:noise_residual} is the \emph{averaged} ASD estimated using Welch's method over 256 second of data in total and 8 second per estimation segment. Therefore the fluctuations in Fig.~\ref{fig:noise_residual} is due to the long-term variation of the RMS of the spot motion. We also show in Fig.~\ref{fig:sample_timeseries} directly the time series to compare the original (blue) and the noise-subtracted (grey) series. In the simulated O3 data, the band-limited RMS in the [20,60]\,Hz varies on the timescale of 10\,s as indicated by the envelopes of the time series. This is because the spot position on the test masses $x_{\rm spot}^{\rm (AC)}$ moves due to the microseismic motion in the $0.1-0.3\,{\rm Hz}$ band. Such a modulation prohibits the removal of the noise using standard signal processing techniques (such as Wiener filter) assuming a stationary coupling. The ``CNN-noise'', nevertheless, successfully mitigates the 10-second-timescale nonstationariety in the time series. 

Furthermore, as we shown in Fig.~\ref{fig:BNS_cumSNR}, with the cleaned sensitivity represented by the grey traces in Figs.~\ref{fig:noise_residual} and \ref{fig:sample_timeseries}, we can get more than $10\%$ of the total SNR 100 seconds prior to the merger. This is sufficient for us to detect nearby BNS events like GW170817 (Sec.~\ref{sec:results}).  For future convenience, we also show the 5- and 95-percentiles at each frequency bin of the residual in the two brown traces in Fig.~\ref{fig:noise_residual}. We further define $\tilde{\rho}$ as the SNR computed assuming a stationary noise background whose values are fixed at the 5-percentiles [i.e., using the lower brown trace for $\sqrt{S}_n$ in Eq.~(\ref{eq:snr})]. We will use this as an estimation of the SNR of the signal at a given distance, though one should keep in mind that $\tilde{\rho}$ will in general be greater than the true SNR of each injection. 

\subsection{Signal detection and classification}
\label{sec:sig_class}

\begin{table}
\caption{Network structure for signal classification. The network includes about 32,000 trainable parameters in total.\label{tab:CNN_class}}
\begin{ruledtabular}
\begin{tabular}{cccc}
\textrm{layer}&
\textrm{output dimension}&\textrm{kernel or}&\textrm{activation}
\\
& & \textrm{pooling size} & \\
\colrule
Conv1D & 64 & 64 & ELU \\
AveragePooling1D & -- & 4 & -- \\
Conv1D & 16 & 16 & ELU \\
MaxPooling1D & -- & 4 & -- \\
Conv1D & 32 & 8 & ELU \\
MaxPooling1D & -- & 8 & -- \\
Conv1D & 16 & 8 & ELU \\
MaxPooling1D & -- & 8 & -- \\
Conv1D & 8 & 4 & ELU \\
MaxPooling1D & -- & 8 & -- \\
Flatten & -- & -- & -- \\
Dense & 16 & -- & ELU \\
Dense & 64 & -- & ELU \\
Dense & 8 & -- & ELU \\
Dense & 4 & -- & Softmax \\
\end{tabular}
\end{ruledtabular}
\end{table}

Once we have trained the CNN-class sub-network, we then inject GW signal onto the cleaned noise background and train the CNN-class for signal detection and classification.

Examples of the input time series to CNN-class is shown in Fig.~\ref{fig:sample_timeseries}. It is the sum of a GW signal at most 256 seconds long (or zero for a null event) and a 256-second residual noise background produced by subtracting the prediction of CNN-noise and the simulated O3 detector noise (i.e., it corresponds to the grey trace in Fig.~\ref{fig:sample_timeseries}).

The training target is the label of each sample: we use (0, 1, 2, 3) for (``Null'', ``BNS'', ``NSBH'', ``BBH''), respectively. We further convert the label into the one-hot representation, so that when we use CNN-class for prediction, the numerical value at each digit gives the probability of the input time series belonging to the corresponding signal class. 

Tabel~\ref{tab:CNN_class} shows the structure of the best performing CNN-class we find empirically. It consists of 5 CNN layers, each followed by a pooling layer (for the first one we use average pooling while for the rest maximum pooling is used). While ReLU activation are used in previous studies, we nonetheless find ELU activation gives a better convergence and therefore we use it for all the CNN layers. We include 3 Dense layers with ELU activation afterwards, and lastly, the output is produced by a Dense layer with the Softmax activation. The sparse categorical crossentropy loss is used together with an Adamax optimizer.

To help the convergence of the network, we utilize the  ``curriculum learning'' approach~\cite{George:18a, George:18b}. That is, we first train CNN-class on very loud GW events with high SNR to guide the NN to an initial convergence. Then we gradually reduce the SNR of injected GW events in the training set to cover the more realistic SNR space of potential astrophysical events. 

Specifically, in the first step, we sample GW events from $\tilde{\rho}(f<f_{\rm cut}^{\rm up})\in [16, 40)$ and with a probability $\propto \left[\tilde{\rho}(f<f_{\rm cut}^{\rm up})\right]^{-2}$, where $\tilde{\rho}(f<f_{\rm cut}^{\rm up})$ is the SNR computed using the 5-percentile noise residual (the lower brown trace in Fig.~\ref{fig:noise_residual}) and integrated to $f_{\rm cut}^{\rm up} = (25, 32, 40)\,{\rm Hz}$ for (``BNS'', ``NSBH'', ``BBH''). 
Note $\tilde{\rho}$ in general will be greater than the true SNR of an injected event because both $f_{\rm cut}<f_{\rm cut}^{\rm up}$ for each realization of the GW event and the background noise is typically greater than 5-percentile value. 
The training set includes $\sim 2,000$ samples for each signal class, plus $\sim 2,000$ samples for null events. Additional 64 samples per class are used as validation. 

Once the first step converges (both the training and loss plateau), we then reduce the SNR range to $\tilde{\rho}(f<f_{\rm cut}^{\rm up})\in [10, 40)$ and $\tilde{\rho}(f<f_{\rm cut}^{\rm up})\in [8, 28)$ in the second and third training steps. In each step, we use $\sim 8,000$ samples per class. There exists a trade-off that training the network to identify low-SNR events would typically degrade its ability to classify null events (i.e., increasing the false alarm rate, or FAR). Consequently, we instead sample events uniformly in SNR in the second and third steps, and do not further lower the SNR of the training data. 

As a comparison, we also construct a network with the same structure as CNN-class but train it on GW time series with stationary noise background generated according to the fundamental O3 sensitivity (which is similar to the aLIGO design sensitivity for $f\lesssim 40\,{\rm Hz}$ of interest; red trace in Fig.~\ref{fig:ASDs_comp}). This reference network is trained with the same curriculum  training steps as CNN-class.

\subsection{Combined network}
\label{sec:combined_prob}

While it is sufficient to train CNN-noise and CNN-class individually as in Secs.~\ref{sec:noise_sub} and \ref{sec:sig_class}, we may further optimize the performance by combining the two networks and training globally. This is because CNN-noise is trained to achieve a broadband noise reduction so that the residual detector noise could potentially serve as the input for pipelines of various purposes. By combining it with CNN-class, the noise subtraction is then optimized specifically for the early detection and classification of GW events. 

To achieve so, we utilize the structure shown in Fig.~\ref{fig:compound_CNN} and load the network weights obtained from individual training as the initial condition for the compound network. We generate $\sim 10,000$ samples for each class with the SNR $\tilde{\rho}(f<f_{\rm cut}^{\rm up})$ uniformly sampled from $[8, 28)$. Each time series of the main GW channel is input to the compound network together with 20 auxiliary channels to internally mitigate the detector noise. The training target, loss function, and optimizer are the same as described in Sec.~\ref{sec:sig_class}. 

We find the compound network could achieve an enhanced performance compared to CNN-class alone (see the discussions in the following section). We will then use the compound network as our final NN and examine its performance of BNS early warning.

\section{Results}
\label{sec:results}

\begin{figure}
  \centering
  \includegraphics[width=0.9\columnwidth]{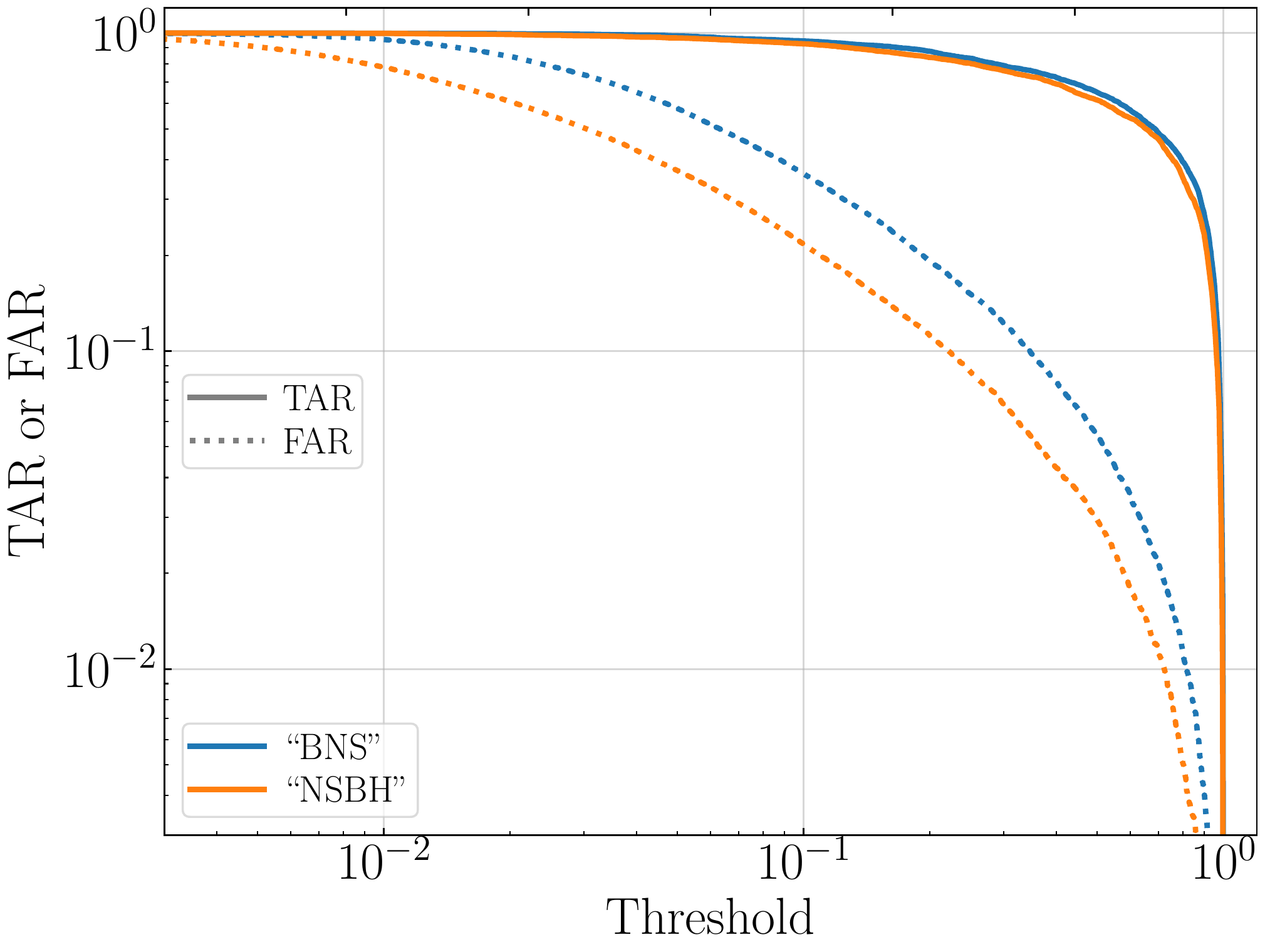}
  \caption{TAR (solid traces) or FAR (dotted traces) as a function of the threshold above which a detection is claimed. The results are obtained using our compound NN (Fig.~\ref{fig:compound_CNN}) with simulated O3 sensitivity. The blue traces is for a typical BNS event at $d_{\rm eff}=40\,{\rm Mpc}$ (see Figs.~\ref{fig:ROC_tar_vs_far} and \ref{fig:det_sen_curve}) and the orange trace is for a NSBH event at $d_{\rm eff}=160\,{\rm Mpc}$ (see Fig.~\ref{fig:det_sen_curve_NSBH}). }
\label{fig:tar_far_vs_thres}
\end{figure}

\begin{figure}
  \centering
  \includegraphics[width=0.9\columnwidth]{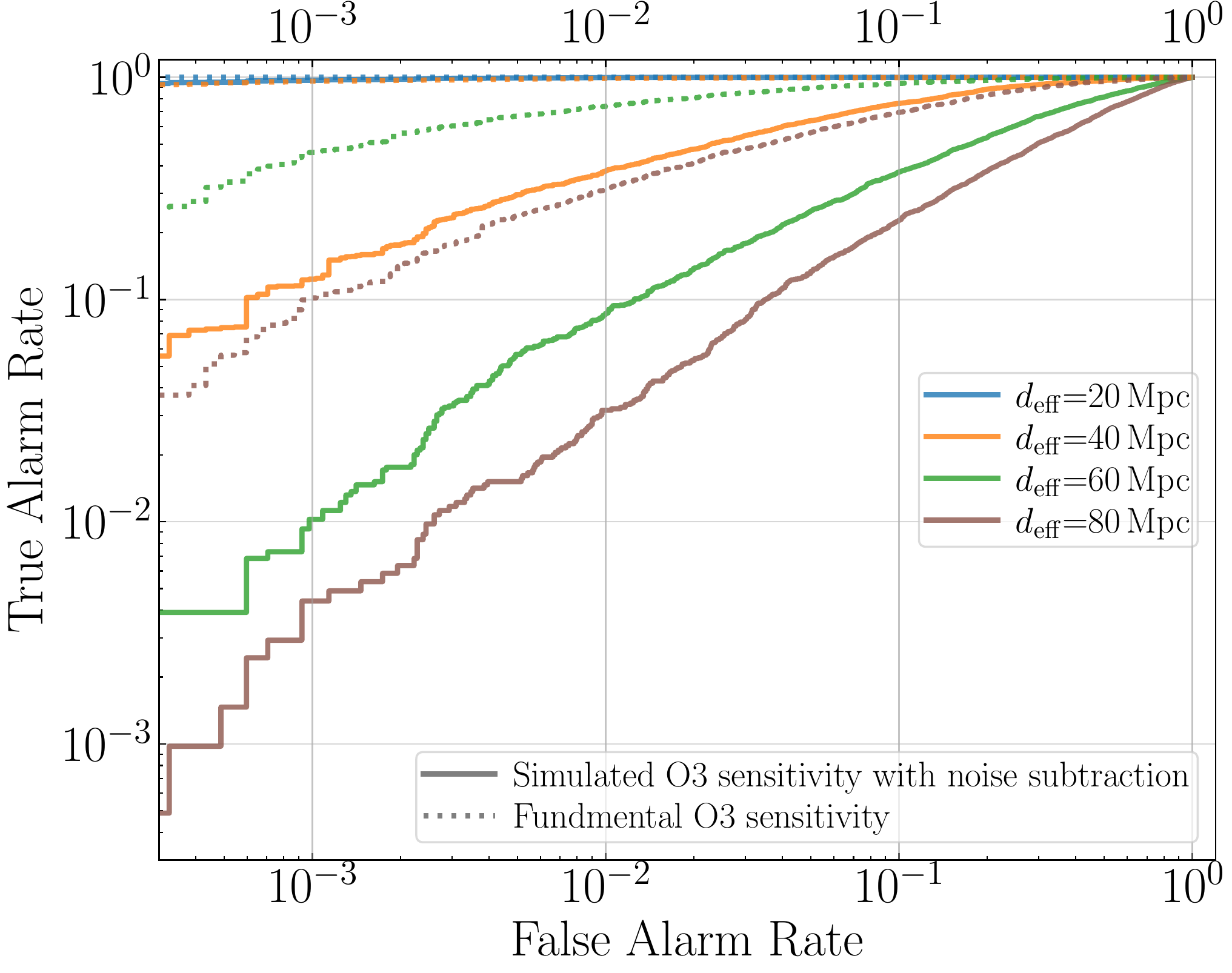}
  \caption{The ROC curves for typical BNS events at different distances. 
  Here we focus on events with $M_1=M_2=1.4\,M_\odot$ and $f_{\rm cut}=25\,{\rm Hz}$ with varying effective distance (assuming $N_{\rm det}=3$). The FAR is constructed from ``null'' events (i.e., detector noise) and ${\rm FAR}=0.01$ corresponds to 1 false alarm every 100 256-second samples (approximately 1 in every 7.1 hours). In the solid traces, we use simulated, non-stationary detector noise representing the O3 sensitivity and the GW readout is input to the compound network together with 20 auxiliary witness channels. As a comparison, the dotted traces use stationary detector noise representing the fundamental sensitivity for O3. }
\label{fig:ROC_tar_vs_far}
\end{figure}

\begin{figure}
  \centering
  \includegraphics[width=0.9\columnwidth]{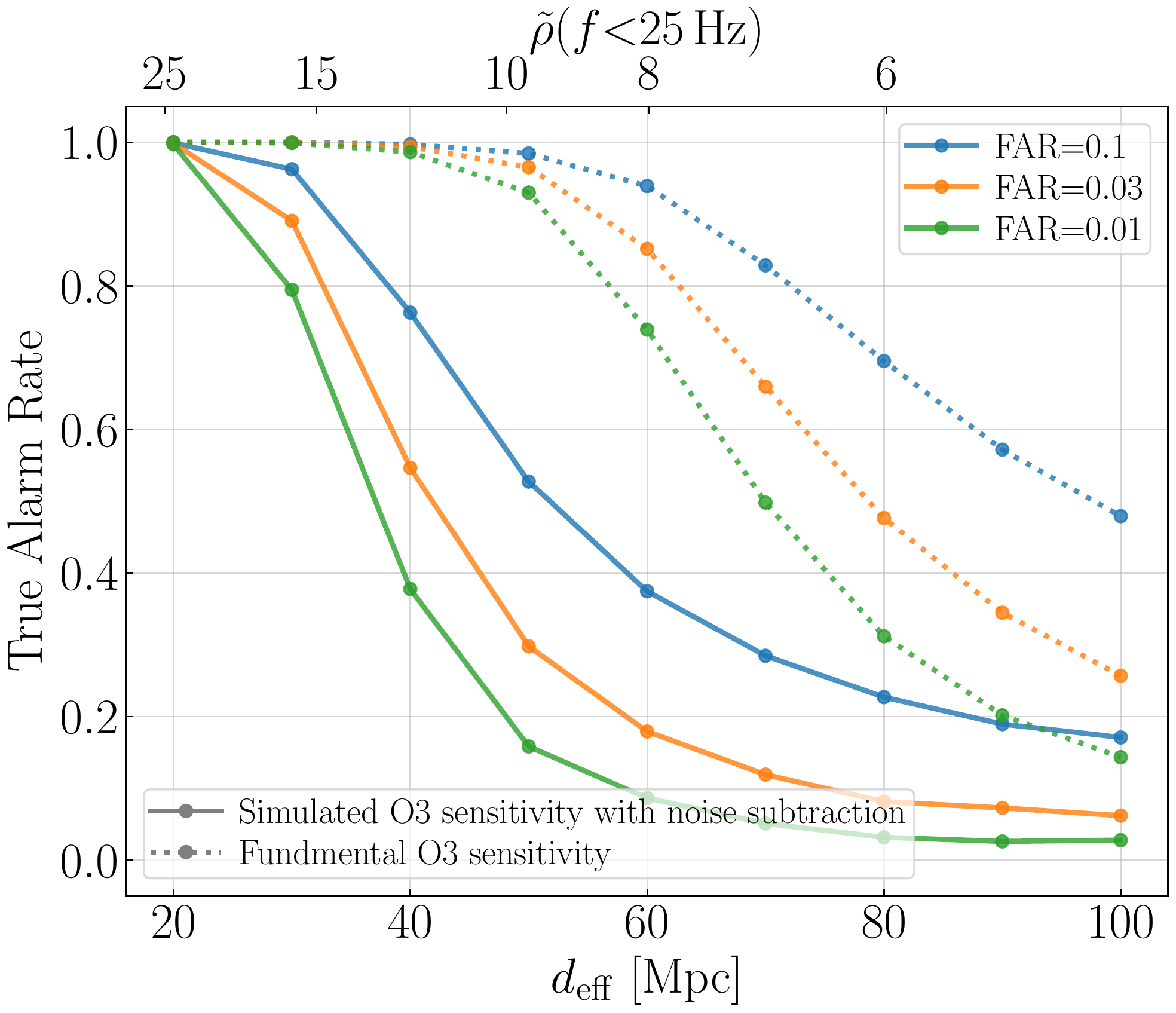}
  \caption{The CNN's detection efficiency of typical BNS events at fixed values of FAR. As a reference, the top x-axis shows the SNR computed using the 5-percentile noise residual (the lower brown trace in Fig.~\ref{fig:noise_residual}; it is typically greater than the true SNR due to both the short and long timescale nonstationarities). A GW170817-like event can be detected about 1.5 minutes prior to the merger if the contamination to the detector's low-frequency sensitivity can be mitigated to the level shown in Figs.~\ref{fig:noise_residual} and \ref{fig:sample_timeseries}. If the design sensitivity is reached, such an early warning would be possible for sources with $d_{\rm eff}\simeq 80\,{\rm Mpc}$.}
\label{fig:det_sen_curve}
\end{figure}

We access the performance of our NN by examining the receiver operator characteristic (ROC) curves which we construct using the \texttt{Scikit-learn} package~\cite{Pedregosa:11}. 
This can be obtained by varying the detection threshold of the predicted true probability and compute both the true alarm rate (TAR) and FAR at the given threshold, as demonstrated in Fig.~\ref{fig:tar_far_vs_thres}.

More conveniently, we can directly consider TAR as a function of FAR for a particular source, as shown in Fig.~\ref{fig:ROC_tar_vs_far}.  For the GW event, we consider BNSs with $M_1=M_2=1.4\,M_\odot$ and $f_{\rm cut}=25\,{\rm Hz}$ and vary the sources' averaged effective distance from $20\,{\rm Mpc}$ to $100\,{\rm Mpc}$ (corresponding to traces of different colors). At each distance, we inject the signal onto 2,000 realizations of the detector noises. The solid traces are results using simulated O3 sensitivity as the noise background with noise mitigation preformed by inputting the auxiliary channels to the compound CNN (Fig.~\ref{fig:compound_CNN}). As a comparison, we also show the performance of the reference network in the dotted traces. It has the same structure as CNN-class but the noise background for training and prediction is generated according to the stationary fundamental O3 sensitivity. Here the FAR is constructed from $\sim 20,000$ realization of detector noises (``null'' events; corresponding to 2 months of data). Note here the rate is measured per 256-second data segment, and as a result, ${\rm FAR}=0.01$ would correspond to approximately 1 false alarm every 7.1\,hr of detector data.

Alternatively, we can fix the FAR and examine how the TAR varies as a function of the averaged effective distance $d_{\rm eff}$. The result is shown in Fig.~\ref{fig:det_sen_curve}. The astrophysical source is still fixed to be BNSs with $M_1=M_2=1.4\,M_\odot$ and $f_{\rm cut}=25\,{\rm Hz}$ and the line styles have the same meaning as in Fig.~\ref{fig:ROC_tar_vs_far}. Different colors now represent different FAR threshold [${\rm FAR}=(0.1, 0.03, 0.01)$ corresponds to 1 false alarm every $(0.7, 2.4, 7.1)\,{\rm hr}$]. We see that if we could mitigate the noise to a level comparable to the grey stripe in Fig.~\ref{fig:noise_residual}, then a GW170817-like event at $d\simeq 40\,{\rm Mpc}$ can be detected 1.5 minutes prior to the merger with a decent chance. 
Because of the nonstationarity in the background noise, the matched-filter SNR is not a constant even for a fixed effective distance. If we nonetheless treat the noise PSD as being stationary and use the 5 and 95 percentiles in the cleaned spectra (i.e., the two brown traces in Fig.~\ref{fig:noise_residual}), we estimate the SNR to be around 12 to 7.3.
On the other hand, if the noise background becomes truly stationary and reaching the designed aLIGO sensitivity, then the early detection can be achieved to $d_{\rm eff}\simeq 80\,{\rm Mpc}$. The corresponding matched-filter SNR is 12. 
The required SNR for detection of a stationary noise background being similar to the SNR calculated using the 5-percentile of the nonstationary background suggests that our final, global training (Sec.~\ref{sec:combined_prob}) mitigates the nonstationarity further and improves the NN's performance than treating the noise subtraction and signal detection as two separate, independent problems.


\begin{figure}
  \centering
  \includegraphics[width=0.9\columnwidth]{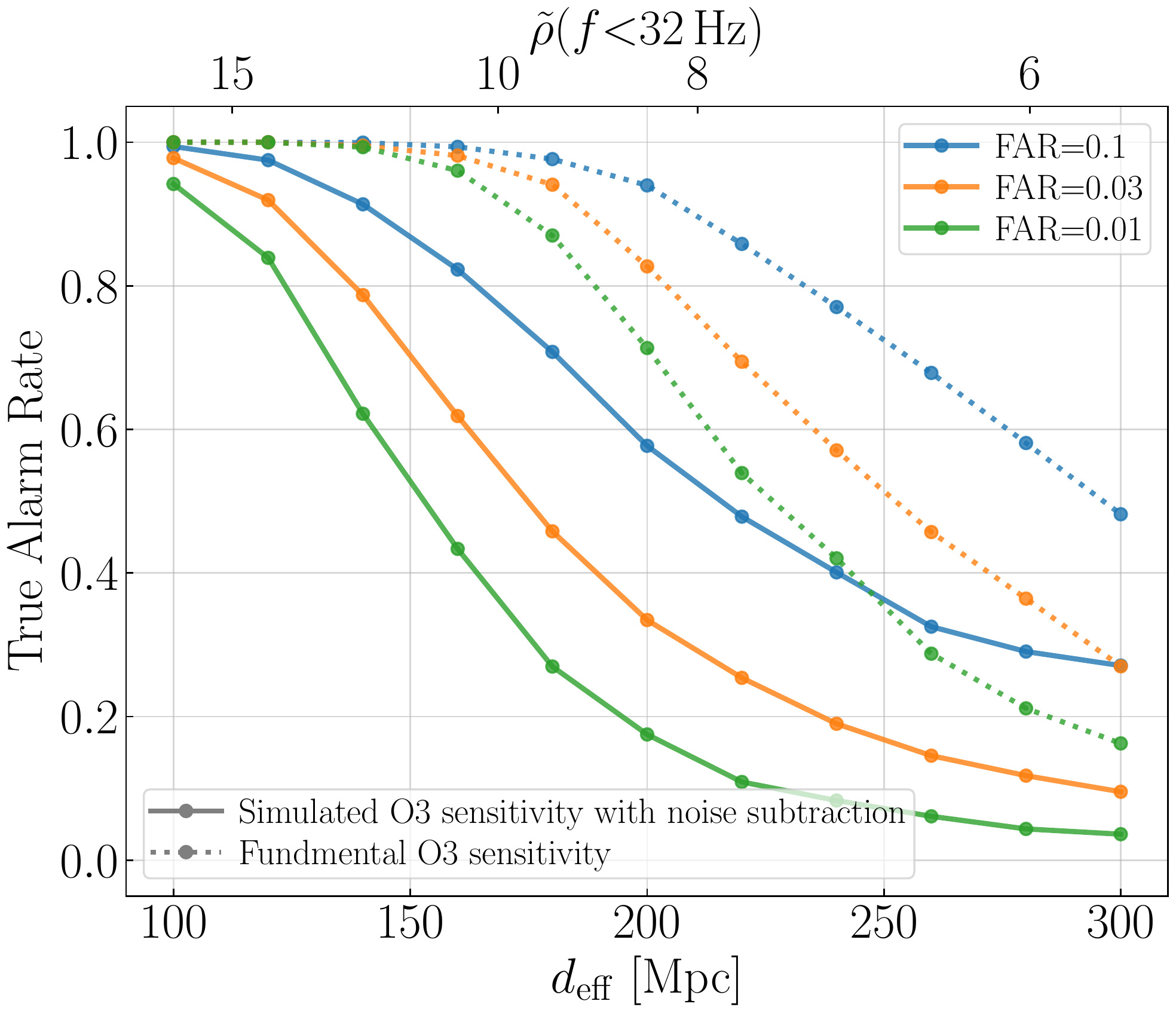}
  \caption{Similar to Fig.~\ref{fig:det_sen_curve} but for an NSBH event with $(M_1, M_2)=(8, 1.4)\,M_\odot$ and $f_{\rm cut}=32\,{\rm Hz}$, with an early-warning time of $t_{\rm m}=12\,{\rm s}$. At a given $\tilde{\rho}$, our NN preforms better for detecting NSBHs than BNSs in general.}
\label{fig:det_sen_curve_NSBH}
\end{figure}

In addition to BNS mergers, NSBH mergers are another type of sources for multi-messenger astronomy, and we access the performance of our CNN for detecting them in Fig.~\ref{fig:det_sen_curve_NSBH}. The GW events we inject are $(M_1, M_2)=(8, 1.4)\,M_\odot$ and $f_{\rm cut}=32\,{\rm Hz}$ ($t_{\rm m}=12\,{\rm s}$). 
{
If we still use the (TAR, FAR)=(0.4, 0.01) as the threshold for detection, we find an NSBH can be detected 12\,s before the merger at an averaged effective distance of $d_{\rm eff}\simeq 160\,{\rm Mpc}$ using simulated O3 sensitivity with noise subtraction. The matched-filter SNR is estimated to be between 10.5 (5-percentile) and 7.0 (95-percentile). Using the stationary, fundamental O3 sensitivity, we find the detection range to be around $d_{\rm eff}\sim 240\,{\rm Mpc}$. The corresponding SNR is 10.5, again similar to the 5-percentile value when the nonstationary noise is used, suggesting that the nonstationarity is largely removed with the internal noise cleaning. 
}

Interestingly, we note that at a given value of $\tilde{\rho}$, our NN typically preforms better for detecting NSBHs than BNSs. Whereas the CNN's sensitivity starts to drop sharply at $\tilde{\rho}\simeq 15$ and essentially vanishes $\tilde{\rho}\simeq 10$ for the ``BNS'' signal, for ``NSBH'' we still have a decent sensitivity at $\tilde{\rho}\simeq 10$. In part, an NSBH has its signal ``concentrated'' in a shorter duration with a louder time-domain amplitude than a BNS and therefore it is more easily recognized by an NN (similarly, Ref.~\cite{Krastev:20} also found that an NN typically performs better for BBHs than BNSs with the same matched-filter SNR). Meanwhile, we have also chosen a higher upper cutoff frequency for NSBHs (32\,Hz) than for BNSs (25\,Hz), and the fluctuation in the PSD of the background noise is less at higher frequencies between different realizations after the cleaning by CNN-noise. 

\begin{figure}
  \centering
  \includegraphics[width=0.9\columnwidth]{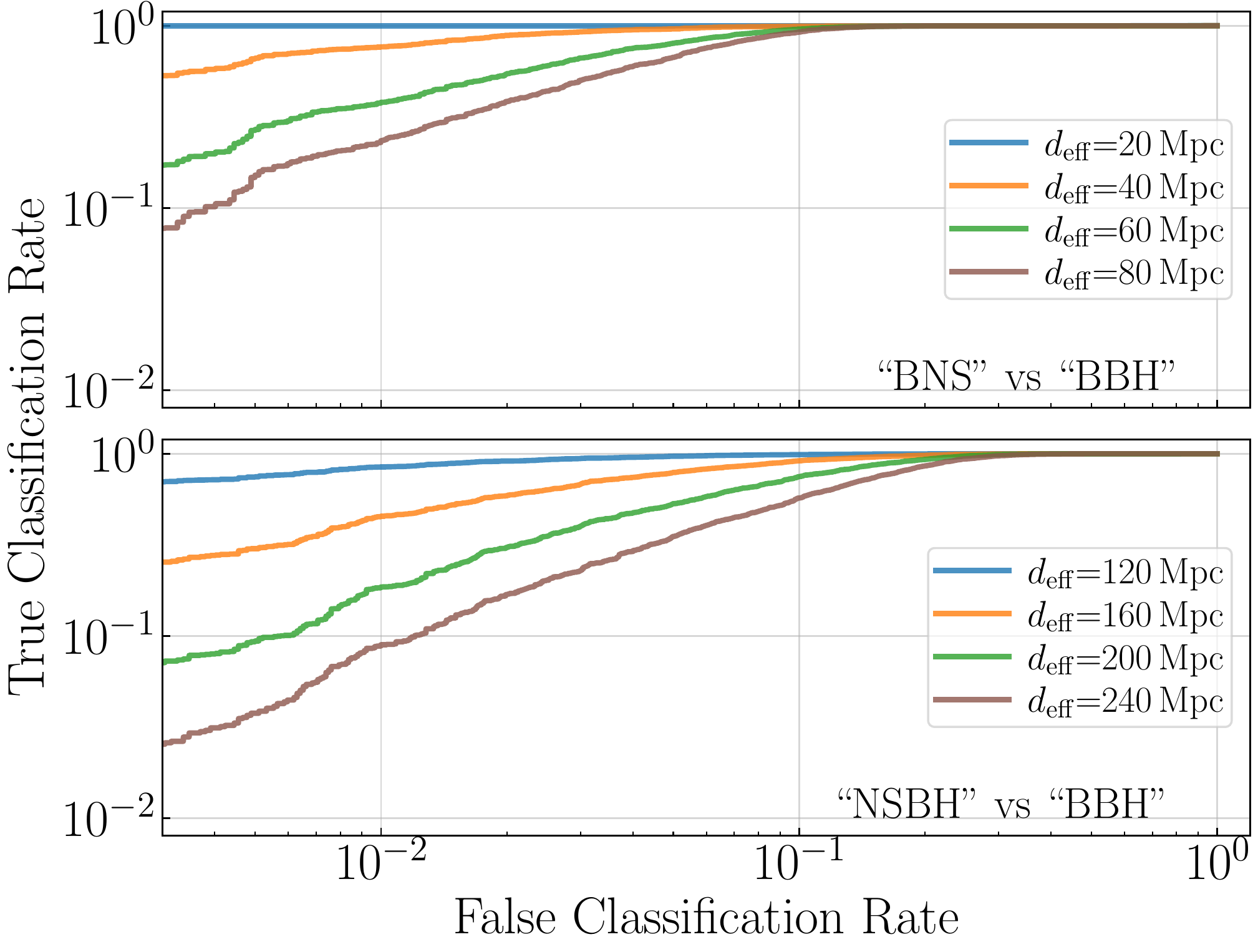}
  \caption{The true classification rate (TCR) as a function of the false classification rate [FCR; i.e., the rate of falsely classifying a ``BBH'' event as a  ``BNS'' (top panel) or an ``NSBH'' (bottom panel) event]. }
\label{fig:ROC_tir_vs_fir}
\end{figure}

Another quantity of interest is the false classification rate (FCR). Specifically, if there is a BBH event (which typically does not have an EM counterpart) present in the GW readout, we want to address the probability of  classifying it as a ``BNS'' or ``NSBH'' and falsely triggering subsequent EM followup observations. The result is shown in Fig.~\ref{fig:ROC_tir_vs_fir}. The FCR is constructed from 5,000 ``BBH'' injections. The ``BBH'' events are sampled from a distribution $\propto \left[\tilde{\rho}(f<40\,{\rm Hz})\right]^{-2}$ and  $\tilde{\rho}(f<40\,{\rm Hz})\in [8, 40)$. By comparing the top panel of Fig.~\ref{fig:ROC_tir_vs_fir} with Fig.~\ref{fig:ROC_tar_vs_far}, we see that a ``BNS'' trigger is much less likely to be confused by a true ``BBH'' event than by the detector noise. The ``NSBH'' class has slightly more false classifications from the ``BBH'' class, yet at ${\rm FCR}=0.01$, we still have ${\rm TCR}>0.1$ for $\tilde{\rho}(f<32\,{\rm Hz})>7$. 


\begin{figure}
  \centering
  \includegraphics[width=0.9\columnwidth]{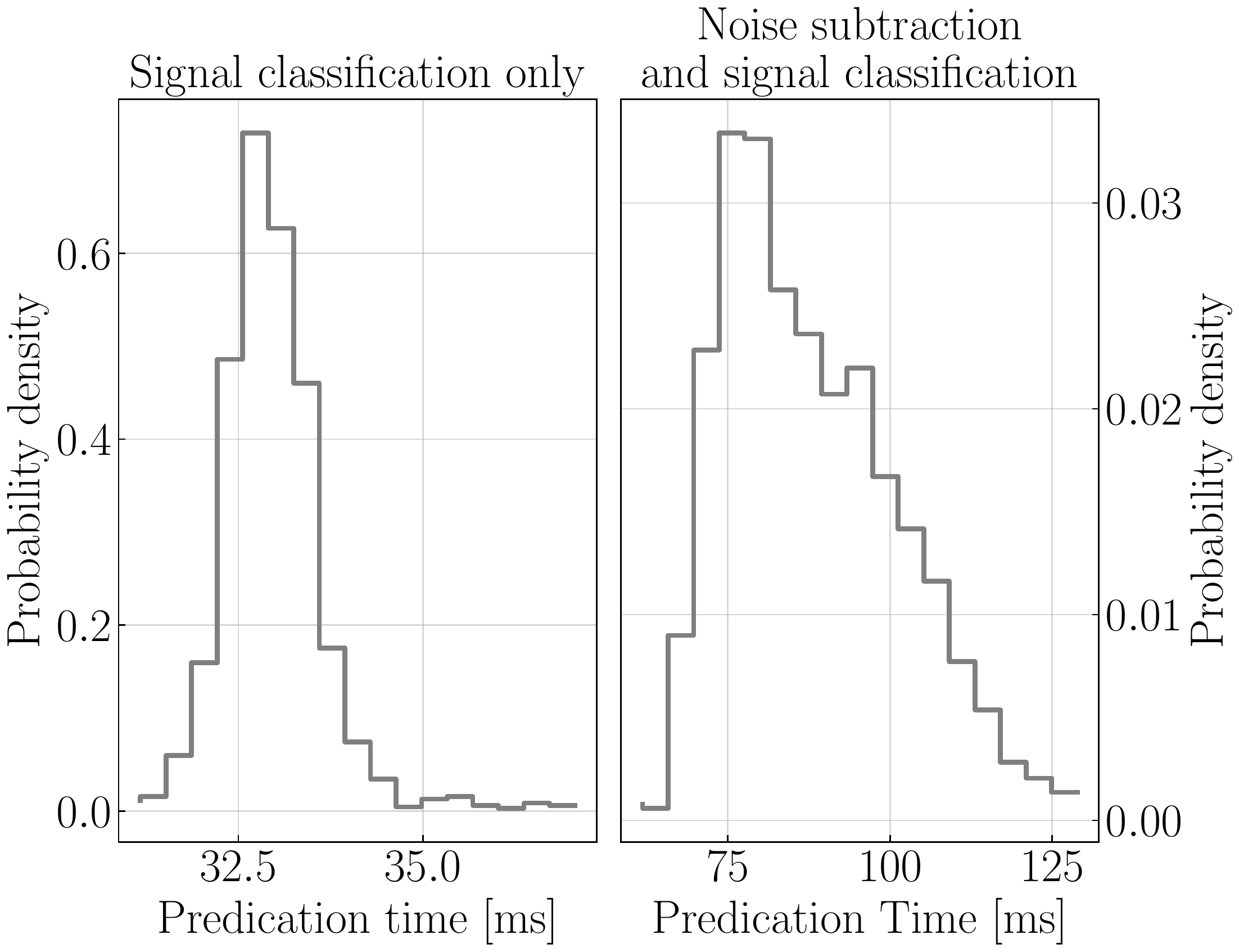}
  \caption{Histogram of the computational time of our CNNs (using a Tesla P100-PCIE-16GB GPU).
  In the left is the time for ``CNN-class'' takes to classify a 256-second time-series, and in the right is the time for the compound problem where we input both the GW readout and 20 auxiliary channels. 
  Once the training is complete, the prediction takes only $\sim 30\,{\rm ms}$ for the NN to classify a time series; even including real-time noise subtraction, the computation time is still less than $100 \,{\rm ms}$ in most cases.}
\label{fig:t_pred_dist}
\end{figure}

Lastly, we point out that our compound CNN not only provides a potential way to achieve real-time noise mitigation and signal detection, it could also serve as an efficient first step to existing match-filter-based pipelines. This is because the computationally expensive part is the training. Once the network is trained, the prediction time is typically only 100 ms for doing both noise mitigation and signal classification, or 30 ms for just preforming signal classification, as shown in Fig.~\ref{fig:t_pred_dist}. Indeed, once a signal is detected and classified by the network, subsequent matched-filter analysis would only need to perform searches over a small sub-bank after the classification preformed by CNN, potentially enhance the efficiency of the existing pipelines. 

\section{Conclusion and Discussion}
\label{sec:conclusion}
We showed that it would be possible to detect BNS (NSBH) signals from the real-time LIGO data series using a ML NN.  

To achieve so, it requires improving the LIGO sensitivity in the $\lesssim60\,{\rm Hz}$ band, which currently dominated by \emph{nonlinear} cross-couplings from the auxiliary control loops and/or environmental perturbations. We demonstrated that one potential way to enhance the low-frequency sensitivity is to input the auxiliary channels together with the main GW readout to an NN and use it to simultaneously preform noise cleaning and signal detection. 

With noise mitigation reaching the level shown in Figs.~\ref{fig:noise_residual} and \ref{fig:sample_timeseries}, we can detect BNS (NSBH) $\sim 100\,{\rm s}$ $(10\,{\rm s})$ prior to merger out to $d_{\rm eff}\simeq 40\,{\rm Mpc}$ $(160\,{\rm Mpc})$ with a ${\rm TAR}\gtrsim 0.4$ and ${\rm FAR}=0.01$ (i.e., 1 false alarm every 7.1 hours). If we have a stationary, Gaussian noise background reaching the designed sensitivity, the early warning can be achieved out to $d_{\rm eff}\simeq 80\,{\rm Mpc}$ and $240\,{\rm Mpc}$ for BNS and NSBH, respectively. 
The matched-filter SNR is 12 and 10 for typical BNSs and NSBHs, respectively. Moreover, we find the threshold SNRs for the Gaussian noise background are similar to the SNRs estimated using the 5-percentile of the nonstationary noise (the bottom brown trace in Fig.~\ref{fig:noise_residual}). This indicates that our compound network structure (Fig.~\ref{fig:compound_CNN}) largely mitigates complications due to a nonstationary background, and the global training (Sec.~\ref{sec:combined_prob}) enhances the NN's performance than treating the noise cleaning and signal detection as two separate problems.

We note that our current NN has not yet reached a sensitivity comparable to the existing low-latency pipelines. For example, Ref.~\cite{Sachdev:20} considered a similar early-warning problem using \texttt{GstLAL} and the designed aLIGO sensitivity. According to the associated data release~\footnote{\url{https://gstlal.docs.ligo.org/ewgw-data-release/data-full.html}}, the authors of Ref.~\cite{Sachdev:20} preformed 1,446 BNS injections with distance from 80\,Mpc to 100\,Mpc in total and they were able to detect 446 (or $31\%$) out of them at an upper cutoff frequency of 29\,Hz and a ${\rm FAR}=(30\,{\rm days})^{-1}$. From the dotted traces in Fig.~\ref{fig:det_sen_curve}, our NN can achieve a similar TAR=0.3 only at ${\rm FAR}=(7.1\,{\rm hours})^{-1}$, a FAR that is about 100 times higher than the \texttt{GstLAL} results. While in part the difference in the performance is due to the fact that we considered a lower cutoff frequency of 25\,Hz and the integration time of the signal is thus $\simeq 30\,{\rm s}$ shorter, it nonetheless indicates that the ML NN still has a large room for future improvement. 

Nevertheless, a ML-based NN has a few advantages over the existing pipelines that warrent it future studying. First of all, as multiple authors have pointed out (see, e.g., Refs.~\cite{George:18a, George:18b, Krastev:20}), an NN is highly efficient in prediction. Indeed, as we showed in Fig.~\ref{fig:t_pred_dist}, it takes the CNN-class only 30\,ms to detect and classify a GW signal from a 256-s data segment. In comparison, the typical latency is about 6\,s for \texttt{GstLAL}, indicating the possibility of accelerating the existing pipelines even further. 

More importantly, we can input not only the strain readout but also auxiliary channels to the NN to enhance the detection of GW signal. Here we focused specifically on removing the excess and nonstationary contamination to the low-frequency band. In addition to help the early warning of BNSs and NSBHs, mitigating the nonstationarity could also help to reduce the false triggers of heavy BBHs due to the drift of background PSD~\cite{Zackay:2019kkv}. Vetoing and/or mitigating glitches is another thing a NN could help with inputting also auxiliary witnesses~\cite{Essick:20, Cuoco:20}. In principle, one can combine multiple noise mitigation feed-forwards and data quality checks with a signal detection routine into a single NN (potentially with a compound structure) that efficiently enhances LIGO's performance. 

As a proof of concept, we used simulated data to mimic the O3 LIGO sensitivity and our auxiliary witnesses are designed to try to emulate realistic channels in LIGO. There is also a public data release containing 3-hours of outputs from all the major LIGO auxiliary channels available at~\footnote{\url{https://www.gw-openscience.org/auxiliary/GW170814/}}. We encourage interested readers to utilize the NN structures we proposed in this work or original NN structures to help the further improvements of the LIGO sensitivity. 


\section*{ACKNOWLEDGMENTS}
We thank Zachary Mark, Katerina Chatziioannou, Erik Katsavounidis, and Deep Chatterjee for useful comments and discussions. H.Y. is supported by the Sherman Fairchild Foundation. 
The authors gratefully acknowledge the computational
resources provided by the LIGO Laboratory and supported by NSF grants PHY-0757058 and PHY-0823459.

\bibliography{ref.bib}

\end{document}